%% file: APS_MainText.tex
\documentclass[%
 reprint,
nofootinbib,
 amsmath,amssymb,
 aps,
]{revtex4-2}

\usepackage{graphicx}
\usepackage{dcolumn}
\usepackage{bm}
\usepackage{hyperref}
\usepackage{multirow}
\usepackage{rotating}
\usepackage[usestackEOL]{stackengine}
\usepackage{array}
\usepackage{xcolor}
\usepackage[utf8]{inputenc} 
\usepackage[T1]{fontenc}
\usepackage{amssymb}
\usepackage{makecell}

\begin{document}

\preprint{APS/123-QED}
\title{Investigative Study on Preprint Journal Club as an\\Effective Method of Teaching Latest Knowledge in Astronomy}

\author{Daryl Joe D. Santos$^{1}$}%
 \email{daryl$\_$santos@gapp.nthu.edu.tw}
\author{Tomotsugu Goto$^{1}$}
\author{Ting-Yi Lu$^{1}$} 
\author{Simon C.-C. Ho$^{1}$} 
\author{Ting-Wen Wang$^{1}$}
\author{Alvina Y. L. On$^{1,2,3}$}
\author{Tetsuya Hashimoto$^{1,2}$}
\author{Shwu-Ching Young$^{4}$}
 \affiliation{$^{1}$Institute of Astronomy, National Tsing Hua University, No. 101, Section 2, Kuang-Fu Road, Hsinchu City 30013, Taiwan. \\ $^{2}$Centre for Informatics and Computation in Astronomy (CICA), National Tsing Hua University, 101, Section 2. Kuang-Fu Road, Hsinchu, 30013, Taiwan. \\ $^{3}$Mullard Space Science Laboratory, University College London, Holmbury St. Mary, Dorking, Surrey, RH5 6NT, United Kingdom. \\ $^{4}$Institute of Information Systems and Applications, National Tsing Hua University, No. 101, Section 2, Kuang-Fu Road, Hsinchu City 30013, Taiwan.}
 
\date{\today}

\begin{abstract}
As recent advancements in physics and astronomy rewrite textbooks in a very rapid pace, there is a growing need in keeping abreast of the latest knowledge in these fields. Reading preprints is one of the effective ways to do this. However, by having journal clubs where people can read and discuss journals together, the benefits of reading journals become more prevalent. We present an investigative study of understanding the factors that affect the success of preprint journal clubs in astronomy, more commonly known as Astro-ph/Astro-Coffee (hereafter called AC). A survey was disseminated to understand how universities and institutions from different countries implement AC. We interviewed 9 survey respondents and from their responses, and we identified four important factors that make AC successful: commitment (how the organizer and attendees participate in AC), environment (how conducive and comfortable AC is conducted), content (the discussed topics in AC and how they are presented), and objective (the main goal/s of conducting AC). These four factors are shown to correlate with each other. We also present the format of our AC, an elective class which was evaluated during the Spring Semester 2020 (March 2020 - June 2020). Our evaluation with the attendees showed that enrollees (those who are enrolled and are required to present papers regularly) tend to be more committed in attending compared to audiences (those who are not enrolled and are not required to present papers regularly). In addition, participants tend to find papers outside their research field harder to read, which makes introducing and explaining basic knowledge without the assumption of the audience already knowing the topic very important. Finally, we showed an improvement in the weekly number of papers read after attending AC of those who present papers regularly, and a high satisfaction rating of our AC. We summarize the areas of improvement in our AC implementation, and we encourage other institutions to evaluate their own AC in accordance with the four aforementioned factors to assess the effectiveness of their AC in reaching their goals.
\end{abstract}

\maketitle


\section{\label{sec:intro}Introduction}

Physics and astronomy are going into the golden era with the recent completion of many new observational facilities and telescopes. Gravitational waves have been detected as Einstein predicted \cite{Einstein1937, Abbott2016, Granot2017}, and B-mode polarization of the Cosmic Microwave Background (CMB, the first confirmation for cosmic inflation theory) is on the verge of detection \cite{Ade2014a, Ade2014b}. Several world-class telescopes have started large surveys of thousands of square degrees of the sky to try to answer outstanding questions in physics such as Dark Energy and Dark Matter. Furthermore, many new telescopes and facilities are planned to operate soon such as NASA’s 6.5m James Webb Space Telescope, European Extremely Large Telescope, and the Thirty Meter Telescope \citep{Clampin2008, Gilmozzi2007, Sanders2013}. All these are thanks to recent advances in technology.
Using these state-of-the-art facilities and telescopes, discoveries are rewriting physics and astronomy textbooks at an unprecedentedly rapid pace. As more advancements are made, not just in astronomy, but also in many fields in science and technology, there is an increasing demand for digesting new information. 
One of the many ways of keeping abreast of the latest knowledge is through journal clubs. A journal club is defined as an education forum of a group of individuals with the agenda of discussing articles to stay up-to-date with literature \cite{Dwarakanath2000}. Journal clubs were founded in the field of medicine in 1875 \cite{Topf2017}, but other studies think that they may have started earlier than that \cite[e.g.,][]{Linzer1987, Mazal2014}.
Reviewing journals has become a routine in most universities and institutions as a way of learning things that are not offered in the traditional classroom. 
In most astronomy institutes, astronomy (preprint) journal clubs are called \textit{Astro-ph/Astro-Coffee} (henceforth abbreviated as \textit{AC}). The former comes from the contraction of \textit{astrophysics}, and the latter comes from the fact that most of these journal clubs serve coffee during the discussion. 

We aim to provide an in-depth investigative study of preprint journal clubs as an effective avenue of teaching the latest advancements in astronomy. 
In particular, our objectives are to:
\begin{itemize}
    \item identify the factors that govern AC; and
    \item assess the AC implementation in our university (National Tsing Hua University: NTHU, Taiwan)
\end{itemize}
To the best of our knowledge, journal clubs in physics and/or astronomy have not yet been extensively studied before. 
However, physicists and astronomers are known to be heavily reliant on the latest literature, hence a detailed assessment of AC activities as a tool to teach the latest astronomy knowledge is timely and relevant.

\subsection{\label{sec:importance}Reading the Latest Literature in Physics and Astronomy}

The main reason for implementing journal clubs is to learn new knowledge through reading the latest literature. This is true for most physics and astronomy students and scientists as suggested by previous studies \citep{Brown1999a, Jamali2008}. 
Cho \cite{Cho2000} also expressed the notion that astrophysicists have replaced journal reading with regular inspection of the Astrophysics subject (\textit{astro-ph}) at \textit{arXiv} (\url{www.arxiv.org}). 
Physicists and astronomers have played a significant role in the establishment of effective scholarly communication and publishing. This was made possible through innovations in scholarly communication methods, since working in physics and astronomy usually require diverse collaborations \cite{Jamali2008}. As a result, they became heavy users of e-print archives \cite{Kling2000, Fry2003}. 

\subsection{\label{sec:preprint}Preprints}

Lawal \cite{Lawal2002} explained that (theoretical) physicists perceive the most recently published information to be the most important. This makes preprints important for physical scientists. By definition, a preprint is an unpublished manuscript meant to be published in journals and/or serials. Preprints can be reviewed/accepted manuscripts, submitted manuscripts that are yet to be reviewed, and circulated manuscripts for peer-review before formally submitting them to a journal
\cite{Lim1996}. In this regard, preprint servers, which are repositories of  preprints, were made. They became the sites of learning the latest advancements in different fields because papers/manuscripts that are yet to be published can be first seen here. One of the most widely used preprint servers is \textit{arXiv}. It was first introduced in 1991 in Los Alamos National Laboratory as its electronic archive of physics preprints. Ever since its establishment, \textit{arXiv} became a crucial instrument for scholarly communication in the field of physics and astronomy, and eventually led preprints to become an integral part of most physicists’ works \cite{Brown2001a, Brown2001b}.

Reading preprints have improved journal clubs. This is because by reading papers that are not yet peer-reviewed, participants may provide the authors of the papers direct criticisms before publication. Besides, preprints are usually free-for-all and available online via open-access preprint servers/archives, allowing easy access to these papers \cite{Casadevall2018}. Integrating preprints in this regard can help students develop critical evaluation skills in understanding manuscripts that are not yet peer-reviewed \cite{Avasthi2018}. Finally, due to their accessibility and the participative interaction that they bring, preprint servers/archives are found to be an integral part of literature resource for astronomers and physicists as they provide scholarly exchanges within the field of astrophysics and physics \cite{Marra2018}.

\subsection{\label{sec:studies}Theories and Studies about Implementing and Assessing Journal Clubs}

Journal clubs are believed to be an effective method of cultivating a community of practice (CoP). A CoP is defined to be a site of individuals that share a common interest and regularly interact as a method of learning \citep{Wenger1998a, Newswander2009, Quinn2014}. It is derived from the concept of situated learning, which describes learning in a social context; that is, learning is achieved through social participation and is possible through situating their role as a member of a large community \citep{Quinn2014, Stein1998, Lave1991}. A CoP must satisfy three components: the domain (the shared interest), the community (the people involved in a regular engagement or interaction), and practice (the shared resources and capabilities being learned throughout the interactions) \citep{Wenger1998a, Wenger1998b}. Previous studies have shown that journal clubs possess these three components and are therefore a good representation of a CoP in an academic setting. For instance, a medical journal club composed of doctors and nurses (the community) usually have the goal of reading and discussing the latest medical journal articles and case readings (the shared resources) to learn new treatment methods and keep abreast of the latest literature (the domain) \citep{Chan2015, Quinn2014, Tallman2016}. Journal clubs are also believed to be an effective method of cultivating CoP, especially at the graduate level. They can serve as bridges for students and faculties at various levels to meet and learn from one another \cite{Newswander2009}. Some journal clubs have also utilized social media as a way of extending engagement to a wider audience. This also provides a solution to workplaces and sites where face-to-face journal clubs/CoP are impossible \citep{Chan2015, Lin2015}.

Journal clubs can also be linked with social constructivism, which highlights the importance of social and individual processes combined to form new knowledge \citep{Palincsar1998}. Journal club attendees are considered to be social constructivists as they learn and build new knowledge by engaging with other people, taking into consideration prior knowledge and experiences as they discuss and criticize literature \citep{Nesbitt2014, Lachance2014}. For example, medical journal clubs are shown to treat discussions as a constructivist way of learning, allowing nurses and doctors to collectively evaluate research findings and case studies and relating them to their prior beliefs and knowledge \citep{Nesbitt2014}.

As for implementing journal clubs, most of the published articles studying journal club implementation are in the field of medicine, and many of these articles have found certain points about journal clubs. For instance, journal clubs usually have a common goal, which is to do “critical appraisal”. This refers to “the process of carefully and systematically examining research to judge its trustworthiness, and its value and relevance in a particular context” \cite{Burls2014}. Without this, journal clubs just become an avenue for attendees to practice and improve their presentation skills \cite{Khan1999}. By implementing critical appraisal in journal clubs, attendees tend to improve their critical analytic skills and the volume of articles that they have read \cite{Rosewall2012}.
They also learn new information which they can apply in actual practice, which is a common goal for medical journal clubs \cite{Milinkovic2008, Alguirre1998}. Having informal discussions of new ideas by the participants (i.e., having a “convivial social forum”), is also a benefit evident in journal clubs, giving them a more casual feel compared to colloquia and seminars \cite{Rosewall2012}.

Many articles have suggested a few points that must be taken into account when implementing a journal club. First, the format of journal clubs may induce secondary effects on the environment of the journal club. For example, didactic-like journal clubs (i.e. journal clubs that do not promote discussion) may cause participants to become passive learners \citep{McGlackenByrne2020}. 
In addition, when senior clinicians organize medical journal clubs, they are more likely to dominate the discussion, by providing more experienced points of view. This may unintentionally cause pressure to other attendees who have less experience. Other factors play a crucial role in journal club success: maintaining attendance and engagement, having good and trustworthy resources for articles (e.g. websites, books, past presentations of other people, etc.), and choosing an organizer who can facilitate the discussion \citep{McGlackenByrne2020}.

Journal clubs become successful when clear and concise goals are prepared beforehand, as establishing a common objective for the attendees is always an important pedagogical step \citep{Gottlieb2018}. 
Different journal clubs may have different objectives. Common objectives include (but are not limited to): teaching and learning critical appraisal skills, keeping up with the latest discoveries in the field, and learning new methodologies which can be applied by the attendees in their respective field. 
Other characteristics of a sustainable and effective journal club include mandatory attendance, regular meetings with suitable schedules for everyone, clear short- and long-term purposes, circulating papers to be discussed before sessions, having an organizer that can lead the discussion, and utilization of the internet for storing and disseminating papers \cite{Deenadayalan2008}. Having a safe space where participants can ask “low-level questions”, everyone can share perspectives without stress or fear of being embarrassed, the balance between the needs of those who are still learning and those who are experts already is met are also considered as guiding principles for journal club design \cite{McGlackenByrne2020}.

Various communication technologies may also be introduced to improve the reach and effectiveness of journal clubs. For instance, the usage of Twitter as a platform for implementing journal clubs stemmed from its ability to facilitate quick real-time dialogue between participants. Twitter journal clubs offer greater flexibility in scheduling sessions, and they are also accessible to a wider audience (i.e. the common public) \citep{Topf2017}. In astronomy, many online websites offer help in implementing AC. A few examples would be Astrobites and VoxCharta. The former is considered as the \textit{Readers' Digest} of astronomy journals, and it publishes accessible and short summaries of latest astronomy research papers \cite{Sanders2017}. On the other hand, the latter is self-defined as a ‘clone of arXiv’ that allows people to vote for discussion and comment on papers posted on arXiv, providing a bridge between real-life scholarly discussions and virtual publication of preprints \cite{Marra2018}. \footnote{(although it has been discontinued as of Jan. 2021)} 
Finally, as for how journal clubs are evaluated, the objectives of the journal club should be checked whether they were met or not \cite{Milinkovic2008}. The effectiveness of (medical) journal clubs must also be evaluated by looking into the discussed articles, the practice of critical appraisal and understanding results, and translating evidence into practice \cite{Deenadayalan2008}. 


\section{\label{sec:method}Methodology}
\subsection{\label{sec:resdes}Summary of Mixed Methods Research Design}

The main goal of this study is to identify the factors that should be considered to implement a successful AC. These factors will then be assessed in our own university's AC as an attempt to see if our implementation is effective in teaching the latest knowledge in astronomy or not. To do this, we formulated a mixed-methods sequential exploratory design consisting of two distinct phases: qualitative followed by a triangulated research design with both quantitative and qualitative phases \citep{Creswell2003}. The whole research design is based on the taxonomy development model, a type of exploratory mixed design model that aims to develop a classification or categorization \citep{Creswell2003,PlanoClark2008}. This model is utilized for identifying the factors that influence the success of AC implementation in the qualitative phase. After identifying these factors, they were used to assess AC implementation in our university with quantitative data to be supported by qualitative data, hence the triangulation.

The qualitative phase's objective is to pinpoint the factors needed to implement an effective AC by learning how other institutions implement their AC and their opinions on how to improve AC. First, a questionnaire was disseminated to obtain an initial understanding of how AC is implemented in other institutions and countries. An interview was implemented next for willing respondents to get more information on their answers in the initial questionnaire. The main results of this phase are the main factors needed to achieve a successful AC implementation.

The triangulated phase, on the other hand, is the second phase of our methodology. Its objective is to see how the factors contribute to the success of our AC by evaluating them with our own AC as a benchmark by looking both into quantitative and qualitative perspectives. It also aims to assess and demonstrate how the identified factors in the qualitative phase are perceived by the participants.

\subsection{\label{sec:qualitative}Qualitative Phase: Initial Questionnaire and Follow-up Interview}

\begin{table*}
\caption{\label{tab:elements}Elements of AC that we asked from the respondents in our initial survey}

\begin{tabular}{| >{\centering\arraybackslash}m{4cm} | m{12cm} |}
\hline
\multirow{4}{*}{\textbf{General Information}} & Name, title, and research field of respondent \\
                             & Number and demographics of institution \\
                             & Number and demographics of AC (if there is) \\
                             & Position of organizer (if the respondent is not the organizer)
                             \\
\hline
\multirow{10}{*}{\textbf{Design/Format of AC}} & Schedule (time, day, frequency) \\
                             & Brief description of how discussion of papers is facilitated \\
                             & Presence of financial aid \\
                             & Food and Drinks \\
                             & Advantages/disadvantages of current AC format \\
                             & Goals of AC \\
                             & Perceived  benefits of AC to attendees \\
                             & Websites/archives where participants look for papers \\
                             & Suggestions/comments about their AC format \\
                             & Is it a mandatory requirement for students?  \\
\hline 
\multirow{2}{*}{\textbf{For those without AC}} & Other activities aside from AC (e.g., colloquium, seminars, etc.) \\
                             & Reason(s) for not having AC \\
\hline 
\end{tabular}
\end{table*}

For the qualitative phase, we designed and prepared a questionnaire to learn more about how other institutes in different parts of the world conduct their AC activities. We had three ways of contacting survey respondents for our qualitative phase. First, we searched the internet for universities and institutions that have AC and/or similar journal clubs by entering on the search engine the keywords “astro-ph”, “astro-coffee”, and “astronomy journal club”. Most of the results showed web pages of their AC indicating the contact information of their organizers. We sent the questionnaire to them via email, introducing ourselves and the goal of our study. Second, we posted our questionnaire to a Facebook group of astronomers (with approximately 9.3K members). These two ways were utilized to reach all possible participants internationally despite the COVID-19 pandemic. Lastly, we asked colloquium speakers in our institute during Spring Semester 2020 to answer our survey. Due to the nature of our method, we cannot estimate how many people were invited to answer our survey.

Table~\ref{tab:elements} shows a list of the elements that we aimed to learn. The questionnaire was tailored to accommodate institutes that have AC and do not have AC. First, we asked for the general information of the respondent. If the respondent has an AC in their institute, we asked for the details about the design/format of their AC. 
The survey was disseminated online (via Google Forms). We sent emails containing the survey questionnaire to AC organizers and representatives. In our questionnaire, we also included a section where we asked the respondents to express interest for a follow-up interview (via video chat or email).
For those who do not have AC, they will be prompted to list down other astronomy-related seminars/events that they organize, and most importantly, their reasons for not conducting AC. This will give us a scope to understand why AC is not being organized in their institutes.
The emails were dispatched last November 2019. We opened the survey for around 4 months, and we also asked the respondents to share the survey with their peers who are also organizing AC activities in their respective universities. More details about the questionnaire are presented in Appendix~\ref{appendixA}.

After receiving the responses from the survey, we analyzed them to construct follow-up questions for the video/email/personal interviews with the survey respondents who agreed to have them. This allowed us to deepen our understanding of their responses by asking follow-up questions related to their survey responses. The interviews helped shed light on (but are not limited to) the following points:
\begin{itemize}
    \item Introduction and academic background of the participant
    \item Opinions about the current format of AC in their institution
    \item Motivations of organizing/attending AC
    \item Environment of AC and interaction of participants and organizers
    \item Improvements and recommendations for their own AC
\end{itemize}

Appendix~\ref{appendixB} presents the interview protocol implemented during the interview sessions, which include the most common questions that were asked. Personal interviews were conducted for the respondents who were also colloquium speakers in our institute during Spring Semester 2020. On the other hand, email and video interviews were conducted for respondents who are outside Taiwan. For video and personal interviews, the interviews were primarily conducted by one of this paper’s authors, while the rest of the authors helped in transcribing the interviews. In vivo coding was implemented to preserve the actual meaning and context of what the participants conveyed in their answers \citep{Saldana2013, Charmaz2006}. All related codes were analyzed thematically and condensed into factors.
After this stage, we identify the main factors that influence the success of AC. We want to assess these factors in our own AC in the next phase.

\subsection{\label{sec:implementation}Triangulated Phase: Implementation and Assessment}
For the triangulated phase, we assessed the AC in our university quantitatively with qualitative data as validation and expansion of our quantitative results. First, we discuss how AC was implemented in our university (NTHU) for assessment. AC in NTHU was implemented as an elective class (ASTR600600: Literature Review in Astronomy Research) during Spring Semester 2020 (March 2020 - June 2020). Each session happens 3 times a week (from Tuesday - Thursday), 12:20 pm - 1:20 pm. 

At the beginning of the semester, the objectives of the AC were made clear to the participants, which are:
\begin{itemize}
    \item To help attendees read efficiently (as many papers as they can);
    \item To keep up with the latest discoveries in astronomy; and
    \item To learn from other fields in astronomy.
\end{itemize}

The format of our AC was structured to meet all of these objectives. To achieve the first objective, a group of regular presenters for each day (5-6 people per group) was assigned. For each day, at least 3 presenters should present their chosen papers via a rotation system. Anyone is welcomed to present papers including the professor of the class (TG), when the 3-paper quota was not achieved or when the session has not yet ended. Everyone is encouraged to ask questions after a paper was presented. A total of 10 minutes per person is allotted for presentation and queries. The second objective encourages attendees to read papers from preprint servers (e.g., arXiv, VoxCharta, and/or Astrobites), journals, and the latest news in astronomy, even if they did not have to present a paper. However, all participants who are required to present regularly are also required to vote in VoxCharta, which is monitored by the professor (TG) every session. Lastly, the third objective could be met by accommodating people studying different research fields. On the first day of AC, the professor (TG) advised presenters to present papers by first discussing the objective(s) of the paper, and why the paper is unique and/or important. Other basic concepts about the paper should also be discussed to accommodate diverse research fields. After that, the main results and conclusions can be discussed.

We classify the attendees into three learner types: \textit{Enrollees} (those who enrolled in AC as an elective class and are therefore required to present a paper every other week), \textit{Volunteers} (those who did not enroll in AC as an elective class but were committed to present papers every other week), and \textit{Audiences} (those who did not enroll and present any papers in AC). Only the enrollees were graded based on the number of papers that they have read and presented within the semester. All enrolled students have an initial grade of B+, which becomes A- if they present one paper, A if they present two papers, and A+ if they present three or more papers throughout the semester.

As for the implementation of our AC during the COVID-19 pandemic, no grave disruptions happened during the sessions, since Taiwan has a relatively good response against the pandemic. However, two of the participants joined AC virtually by connecting via Facebook/Google Meets; one went to quarantine right after arriving in Taiwan during the first few months of AC, while the other went back to her home country before the start of the Spring Semester 2020. Due to the university’s rules against COVID-19, we were not allowed to eat and drink in the classroom. This prevented us from eating lunch during AC, which had a timeslot during lunchtime.

We investigated the effectiveness of the format of our AC as an elective class during the Spring Semester 2020 by evaluating the factors via a questionnaire that we disseminated to our AC attendees at the end of the semester. The questionnaire contains Likert-scale questions about the important points for each factor and questions confirming whether our specific objectives were met or not. A short answer question for comments, suggestions, and recommendations for improvement was also included in the questionnaire. This helped us pinpoint which factor needs improvement in our implementation. In addition, it can serve as a support to our quantitative data and gauge how our participants feel about our AC.

\section{\label{sec:results}Results and Discussion}
\subsection{\label{sec:surveyresults}Survey Results}

We were able to interview 9 people who implemented AC in their institutions and responded to our initial survey. In addition, 2 more people (who are both postdocs) who answered our survey mentioned that they did not have AC activities in their institutions. The 9 interviewees' ACs all utilize \texttt{arXiv} and/or VoxCharta to look for preprints to discuss. 
Table~\ref{tab:characteristics} summarizes the basic information of the 9 respondents with who we were able to have a follow-up interview. Based on our in vivo coding, we counted how many respondents discussed the coded themes/points in their responses and considered only the points with at least N=3 responses (one-third of the total number of interviewees). We grouped these themes into four distinct categories which we declare as the four important factors for a successful AC. Table~\ref{tab:factors_summary} summarizes the results of our theming.


\begin{table*}[hbt!]
    \caption{Characteristics of the 9 participants who had a follow-up interview with us.}
    \begin{tabular}{|c|c|c|c|c|}
    \cline{1-5}
    \textbf{Gender} & \textbf{Based on} & \textbf{Academic Position} & \textbf{Research Group/Field} & \textbf{Email, Video, or Personal Interview?} \\ \cline{1-5}
    M & USA & Faculty & N/A & Video \\ \cline{1-5}
    M & USA & Postdoc & Stars & Email \\ \cline{1-5}
    M & USA & Faculty & X-ray Group & Video \\ \cline{1-5}
    F & UK & Graduate & Theoretical Astrophysics & Personal \\ \cline{1-5}
    M & Australia & Graduate & Galaxy Evolution and Formation Group & Personal \\ \cline{1-5}
    M & USA & Graduate & Transient and Variable Science Group & Video \\ \cline{1-5}
    F & Canada & Faculty & N/A & Email \\ \cline{1-5}
    F & USA & Graduate & Radio Astronomy/Pulsars/Fast-radio Bursts & Email \\ \cline{1-5}
    F & USA & Postdoc & Observational Cosmology & Personal \\ \cline{1-5}
    \end{tabular}
    \label{tab:characteristics}
\end{table*}

\subsection{\label{sec:factors}4 Important Factors for a Successful AC}

In this section, we discuss the 4 main factors to implement an effective and successful AC.

\begin{table}[hbt!]
\caption{\label{tab:factors_summary}%
Summary of the four main factors affecting the success of AC which resulted from theming the codes from the interviews. The left column shows the factors, while the middle column shows the main themes/points. The right column shows the number of interviewees who discussed the aforementioned main points (total number of interviewees is 9).}
\begin{ruledtabular}
\begin{tabular}{ccc}
 \textbf{Factor} & \textbf{Main Theme/}Point & \textbf{Number (N)} \\ \hline
 \multirow{4}{*}{Commitment} & Mandatory/For-credit class & 3 \\
                             & Dedication & 3 \\
                             & Importance of organizer & 5 \\
                             & Lack of Commitment & 2* \\ \cline{1-3}
 \multirow{3}{*}{Environment} & Casual & 5 \\
                              & Pressure & 3 \\
                              & Food/Drinks & 7 \\ \cline{1-3}
 \multirow{2}{*}{Content} & Diverse Topics/Platforms & 6 \\\
                          & Bias & 3 \\ \cline{1-3}
 \multirow{2}{*}{Objective} & Importance of Purpose & 3 \\
                            & Different Goals & 4 \\
 
\end{tabular}
\end{ruledtabular}
\begin{flushleft}
*2 other respondents from our initial survey, who are postdocs and do not have AC in their institute (and thus not included in the interviewee list), mentioned that the main reason why they do not have it is due to lack of commitment.
\end{flushleft}
\end{table}

\subsubsection{Commitment}
\label{sec:commitment}
The first factor is commitment, which refers to the dedication of the people involved in AC. This is the factor that focuses on the active participation of the attendees and organizer/s in AC. Some of the interviewees mentioned that having AC as part of the curriculum (i.e., as an elective class where students can get credits, or as a mandatory subject for postgraduate students) can also encourage participation, and instill commitment to the AC participants (N=3). Some of them also emphasized the importance of dedication in attending/organizing AC (N=3). The 3 respondents who mentioned this point highlighted that their AC activities are successful due to having a solid core of students and professors who regularly attend sessions and initiate discussions. A majority of our interviewees also believe in the importance of having a responsible and committed organizer to head the AC (N=5). However, there are mixed opinions as to who should organize the discussion in AC. Among the 5 interviewees who mentioned the importance of organizers, 2 of them personally believe that students may be less effective in organizing AC. This is because students generally have lesser experience in research compared to other senior members, so they may not be able to foster more in-depth discussions. Besides, when faculties do not attend AC, other students might perceive AC as an unimportant event. This makes the presence of faculties, especially in the organizing team, crucial for AC. On the other hand, three of the respondents believe that being an organizer does not depend on any academic status. These respondents have students as their AC organizers, and it may also bring some advantages, For instance, student organizers can use their relationships with other students to invite them to attend, and promote a more relaxed environment where everyone can ask questions of varying difficulty, which others may consider as daunting in front of more senior scientists as organizers. Also, lack of commitment is identified as the main reason for failing to maintain or even starting an AC (N=2). Coincidentally, 2 of our survey respondents who do not have AC in their institution (who we did not interview) both mentioned the lack of commitment in establishing AC as the main reason why they do not have an AC.
An organizer, regardless of their academic position (student, postdoc, or professor), must be committed so that AC activities are implemented properly to reach the main goal of AC. Previous studies \cite{McGlackenByrne2020, Deenadayalan2008, Kleinpell2002} mentioned the importance of finding an appropriate organizer that can maximize the given time for a productive discussion and can meet the balance between selecting an expert who can lead the discussion and a junior who may not know much about the material and can therefore engage with the discussion. The attendees, on the other hand, must be committed to actively participate during the discussion, as well as to cooperate with the organizer to attain the main goal of AC. This can be done in various ways (e.g., by asking and answering questions, posting/voting and presenting papers in every session, etc.). 
Previous works also highlighted the importance of encouraging attendance and participation to create a sustainable and effective journal club, in which mandatory attendance may work well in this case. On the other hand, voluntary participation may lead to greater satisfaction levels of the participants, but at the cost of lower levels of learning \cite{Deenadayalan2008}. Lastly, \citep{Lave1991} also emphasized that the first component of a CoP, the domain or shared interest, implies that all members of the community must be committed in the domain. This further reiterates the importance of commitment in successfully implementing AC.

\subsubsection{Environment}
The second factor, environment, describes the overall façade of how AC is conducted. Most of the respondents mentioned the importance of a casual environment in AC to make sure that it is conducive for active discussion and arguments (N=5). Several interviewees pinpointed some factors that induce pressure for the participants, such as having the presence of postdocs and professors, and the formal presentation of papers, which can cause some participants to be less interested to attend due to pressure (N=3). Some students feel embarrassed when they ask basic questions in front of senior researchers, discouraging them from participating and learning. 
This dilemma was also highlighted by previous studies about medical journal clubs, wherein discussions dominated by experts/senior clinicians may lead to a restricting environment for amateur clinicians to ask questions \cite{McGlackenByrne2020}. This calls for the importance of accessibility for every learner type when the goal is to reach more students. However, some AC activities might be tailored to focus on improving skills in presenting topics to a broad range of audiences and summarizing papers more effectively, making their AC more formal than others. This was the previous AC format for one of our respondents, who stated that after switching to a less formal format, there appeared to have more student participation.

Previous studies \cite{Freshwater2011, Rosewall2012, Topf2017} have mentioned two particular aspects that separate journal clubs (like AC) compared to colloquia and seminars: their ‘collegiality’ and ‘conviviality’. These two aspects make AC a social and casual event. After all, these papers can be presented more formally in a conference or a lecture, but the sociality during AC is what makes it distinct from other astronomy-related seminars \cite{Freshwater2011, Rosewall2012, Topf2017}.
Instead of preparing and presenting slides, participants can casually discuss the paper by e.g., flashing the paper on screen and scrolling to the relevant parts for discussion. The social aspect of AC can also be related to AC participants as social constructivists, constantly engaging in discussions and importing prior knowledge in learning new literature \cite{Nesbitt2014, Lachance2014}. The feeling of being part of a larger community and collaboration also gives participants more opportunities to talk with each other. This helps promote camaraderie and sociality \cite{Weber2008}

Introducing food before, during, or after AC may also help in maintaining the sociality of AC. A majority of our interviewees mentioned that they have food and/or drinks during their AC (N=7). Among these 7 respondents, 2 of them have their coffee provided by their department, allowing everyone in the department to have coffee anytime, including AC time. One of them also suggested that eating lunch while having AC is a good idea, as it saves time. On the other hand, some of them buy foods to share during AC. However, only two of all the respondents believe that food and drinks help in promoting engagement and commitment, while the rest believe that these are not necessary for a successful AC. Previous studies have also shown that providing food helps increase attendance and prolong the longevity of AC implementation \cite{Spillane1998, Stange2006}. Introducing food in AC also reduces formalities, providing a casual environment for discussion \cite{Topf2017}.

\subsubsection{Content}
The content of AC refers to the topics which are discussed in each session and the platforms where materials for discussion are presented and/or come from. Most of our interviewees discuss a variety of topics during AC (N=5). Among these 5 respondents, 1 of them considered AC as a class in their university where some of the students who attend do not have much background yet in astronomy, so they get to learn astronomy in AC. This is why they keep the content of their AC accessible to all possible levels of education. For those who are more experienced, discussing diverse topics works best. Most of the respondents mentioned that they can discuss various topics in AC because they always try to explain jargon in layman terms. Discussing a variety of topics might be hard for some people who are only familiar with their research field, and so having a good introduction about the topic of the paper to be discussed is important to overcome this obstacle. After a few times of introducing the same topic, everyone will be familiarized with the topic and at some point, the introduction will not be needed for the regular attendees. Discussing recent astronomy news and videos of seminars/colloquia can also help participants feel less restricted to just journals/preprints. This shows that learning discoveries in astronomy is not limited to reading papers. Having a good resource for discussion is always important for all journal clubs, and this is not only limited to published journals and preprints. Online “informal” materials such as websites, social media platforms, blogs, and forums\footnote{Examples are ScienceDaily (\url{https://www.sciencedaily.com/}), Astrobites, etc.} may be utilized to discuss controversies and practice critical thinking outside the scope of scholarly environment \cite{McGlackenByrne2020}. Online journal clubs are also proof that journal clubs like AC can also be conducted in various media \citep{Chan2015, Lin2015}.

One particular problem that some ACs face is that participants tend to choose only papers that they like or they understand clearly (e.g., papers related to their research field). We call this topic bias (N=3). This usually happens when the AC is attended only by people who are working in the same research group. However, if the goal of the AC is to learn more astronomy concepts, this must be avoided to expand the field of knowledge to be learned in the AC. This also limits the diversity of the participants as others may perceive AC as a ‘group meeting’ in the same field and not as a multi-disciplinary class. Among the 3 respondents who highlighted topic bias, 1 of them mentioned that one way they minimize topic bias is to encourage presenters in AC to present papers that are not related to their research field, while another mentioned that if everyone has their favorite topics to discuss, having diverse topics in AC will come out naturally. Since most medical journal clubs focus on one particular specialty \cite{Harris2011}, topic bias is usually not present in medical journal clubs.

\subsubsection{Objective}
\begin{figure}[hbt!]
\includegraphics[scale=0.25]{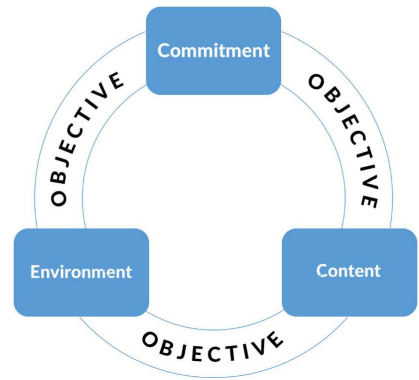}
\caption{\label{fig:factors} Diagram showing the 4 factors needed for implementing a successful AC and how they are connected with each other.}
\end{figure}

The last factor that affects the success of AC is objective. A CoP such as AC must have shared goals which everyone can achieve together. 
The main goals must be clarified first when starting or organizing an AC activity. Various objectives require various elements to keep AC consistent and productive. Therefore, the objective of AC can shape the format of AC implementation (N=3). Some usual objectives of AC which are implemented (and/or recommended) by our interviewees are as follows (but are not limited to):
\begin{itemize}
    \item Introduce different research fields to people from other fields
    \item Learn more about your research field
    \item Help each other’s research by discussing many papers per week
    \item Serve as an astronomy class for people without much astronomy background 
\end{itemize}
For example, among the 3 respondents who mentioned how purpose shapes the format of AC, one of them mentioned that they do AC everyday as a reminder for everyone in their institute to check \textit{arXiv} everyday. Another mentioned that he and his peers started an AC exclusive for students only to lessen the pressure induced by having senior scientists attending AC. 
Besides, having many objectives at once is also possible (N=4), but organizers and participants must work hand-in-hand to achieve these. The objective(s) of AC affects the demographics of its participants (which is somehow related to commitment), the topics discussed (content), and how conducive AC is for learning without any prejudice (environment). One of the respondents purported that having different objectives produce different advantages in AC. For instance, if the goal of the AC is to teach more advanced topics in astronomy that are not usually discussed in classes, then the tendency is that fewer students would be interested to attend as they would think that it is too hard for them to understand. However, attendance is a very dynamic aspect because it is dependent on many factors, such as how AC is advertised, how AC is perceived, and its schedule (time and frequency). These things must also be taken into consideration when gauging the commitment of people who are invited to join AC. Conflicting schedules of participants also give organizers a hard time in finding a common time for AC. This is hard to solve as it happens on a case-by-case basis. 
As what previous literature suggested, establishing the main goal must be the first step in creating a journal club or any educational activity in general as this will be the main basis for reflection and evaluation. In addition, these objectives must be reviewed regularly and approved by all participants and must be explicitly explained to let everyone be aware of them \cite{Gottlieb2018, Mazal2014}. 
Fig.~\ref{fig:factors} shows how these factors are interconnected. The objective serves as a tether that connects the other three factors, namely commitment, environment, and content. This is because once the objective has changed, the way the rest of the factors are implemented will change as well.

\subsection{\label{sec:NTHU}Implementation and Assessment of AC in NTHU}
In this section, we present the statistical analyses we implemented to assess AC in our university, NTHU, and the results of these analyses.

\subsubsection{Statistical Analysis}
Following previous studies \cite{McGlackenByrne2020, Harris2011}, we look into the different learner types (i.e., enrollees, volunteers, audiences) for evaluation. A 5-point Likert scale (with 1 being strongly disagreed/least likely, and 5 being strongly agreed/most likely) was utilized to evaluate certain points related to the aforementioned 4 main factors. 

For the Likert-scale responses, the frequencies are shown in Table IV. As a countercheck, we also employed non-parametric statistical tests to calculate which pair of groups show a significant difference: Kruskal-Wallis test for three groups (with Dunn Test with Bonferroni correction as a post-hoc test), and Mann-Whitney U test for two groups (since some questions did not apply to audiences, hence only having two groups available for comparison). We utilized the Python package \texttt{SciPy} \cite{Virtanen2020} for the aforementioned statistical tests. We caution the readers that we are limited to the number of respondents that we have: 10 Enrollees, 8 Volunteers, and 2 Audiences. Therefore, the statistical power of our tests may not be as robust as desired. Despite this, we still show the results of our statistical tests to quantify possible differences among the groups’ perspective in each factor. These will be helpful for future works which can utilise a larger sample size for the same purpose (see Sec. \ref{sec:limitations}).

We also investigated the preferred astronomy fields that our attendees tend to discuss and fields that they avoid discussing. We asked for the research fields, discussed topics, topics that they find easy to understand (hereafter easy topics), and topics that they find difficult to understand (hereafter difficult topics). The topics that they can choose were based on the list of topics in arXiv's \textit{astro-ph} section. We used the non-parametric Fisher exact test due to its applicability with nominal data and its robustness with small sample sizes, even with having zero sample size in some of the groups \citep{Fisher1956}. By using a two-tailed Fisher exact test, we quantified whether there is a significant difference between the research topics and discussed topics, easy topics, and difficult topics among each participant type. The two-tailed Fisher exact test was computed using the calculator on VassarStats website (\url{http://vassarstats.net}).


\subsubsection{Assessment Results for AC in NTHU}
In this section, we summarize and discuss the results of our Likert-scale questionnaire. Table \ref{tab:freq_dist} shows the frequency table of our responses.


\begin{sidewaystable*}
\caption{\label{tab:freq_dist}Frequency distribution of Likert-scale questionnaire results. The number of respondents per group (No.) and median (Med.) are reported as well.}
\centering
\begin{tabular}{c|c|ccccccc|ccccccc|ccccccc|cccccc} 
\hline
\multirow{2}{*}{Factors }                                        & \multirow{2}{*}{Questions}            & \multicolumn{7}{c|}{Enrollees} & \multicolumn{7}{c|}{Volunteers}                                          & \multicolumn{7}{c|}{Audiences} & \multicolumn{6}{c}{Total}   \\ 
\cline{3-29}
                                                                 &                                       & No. & 1 & 2 & 3 & 4 & 5 & Med. & No. & 1 & 2 & 3 & 4                      & 5                      & Med. & No. & 1 & 2 & 3 & 4 & 5 & Med. & No. & 1 & 2 & 3  & 4  & 5   \\ 
\hline
\multirow{3}{*}{Commitment}                                      & Commitment in attending AC           & 10  & 0 & 0 & 1 & 5 & 4 & 4.0  & 8   & 0 & 0 & 4 & 3                      & 1                      & 3.5  & 2   & 0 & 0 & 2 & 0 & 0 & 3.0  & 20  & 0 & 0 & 7  & 8  & 5   \\
                                                                 & Participation during discussion       & 10  & 0 & 2 & 4 & 3 & 1 & 3.0  & 8   & 0 & 1 & 3 & 4                      & 0                      & 3.5  & 2   & 0 & 0 & 2 & 0 & 0 & 3.0  & 20  & 0 & 3 & 9  & 7  & 1   \\ 
\cline{15-15}
                                                                 & How likely to attend AC next semester & 10  & 1 & 0 & 1 & 3 & 5 & 3.5  & 8   & 0 & 0 & 0 & 2 & 6 & 5.0  & 2   & 0 & 0 & 2 & 0 & 0 & 3.0  & 20  & 1 & 0 & 3  & 5  & 11  \\ 
\hline
\multirow{7}{*}{Environment (Comfort in the following elements)} & Schedule and frequency                & 10  & 0 & 1 & 2 & 3 & 4 & 4.0  & 8   & 0 & 0 & 2 & 2                      & 4                      & 4.5  & 2   & 0 & 0 & 2 & 0 & 0 & 3.0  & 20  & 0 & 1 & 6  & 5  & 8   \\
                                                                 & Attending AC                          & 10  & 0 & 0 & 1 & 5 & 4 & 4.0  & 8   & 0 & 0 & 0 & 4                      & 4                      & 4.5  & 2   & 0 & 0 & 1 & 0 & 1 & 4.0  & 20  & 0 & 0 & 2  & 9  & 9   \\
                                                                 & Presenting                            & 10  & 0 & 1 & 3 & 3 & 3 & 4.0  & 8   & 0 & 0 & 1 & 4                      & 3                      & 4.0  & 0   & 0 & 0 & 0 & 0 & 0 & 0.0  & 18  & 0 & 1 & 4  & 7  & 6   \\
                                                                 & Reading papers before presenting      & 10  & 0 & 1 & 3 & 4 & 2 & 4.0  & 8   & 0 & 0 & 1 & 4                      & 3                      & 4.0  & 0   & 0 & 0 & 0 & 0 & 0 & 0.0  & 18  & 0 & 1 & 4  & 8  & 5   \\
                                                                 & Rotation system                       & 10  & 0 & 0 & 3 & 2 & 5 & 4.5  & 8   & 0 & 0 & 1 & 2                      & 5                      & 5.0  & 0   & 0 & 0 & 0 & 0 & 0 & 0.0  & 18  & 0 & 0 & 4  & 4  & 10  \\
                                                                 & Using VoxCharta                       & 10  & 0 & 0 & 2 & 4 & 4 & 4.0  & 8   & 0 & 2 & 4 & 0                      & 2                      & 3.0  & 0   & 0 & 0 & 0 & 0 & 0 & 0.0  & 18  & 0 & 2 & 6  & 4  & 6   \\
                                                                 & Using Facebook Group                  & 10  & 0 & 1 & 2 & 3 & 4 & 4.0  & 8   & 0 & 0 & 1 & 3                      & 4                      & 4.5  & 0   & 0 & 0 & 0 & 0 & 0 & 0.0  & 18  & 0 & 1 & 3  & 6  & 8   \\ 
\hline
\multirow{5}{*}{Content}                                         & Comfort in papers within their field  & 10  & 0 & 0 & 2 & 3 & 5 & 4.5  & 8   & 0 & 0 & 0 & 3                      & 5                      & 5.0  & 2   & 0 & 0 & 0 & 1 & 1 & 4.5  & 20  & 0 & 0 & 2  & 7  & 11  \\
                                                                 & Comfort in papers outside their field & 10  & 0 & 1 & 7 & 1 & 1 & 3.0  & 8   & 0 & 0 & 2 & 1                      & 5                      & 5.0  & 2   & 0 & 0 & 0 & 1 & 1 & 4.5  & 20  & 0 & 1 & 9  & 3  & 7   \\
                                                                 & Amount of new knowledge gained        & 10  & 0 & 0 & 2 & 6 & 2 & 4.0  & 8   & 0 & 0 & 0 & 6                      & 2                      & 4.0  & 2   & 0 & 0 & 0 & 2 & 0 & 4.0  & 20  & 0 & 0 & 2  & 14 & 4   \\
                                                                 & Is gained knowledge useful?           & 10  & 0 & 1 & 6 & 2 & 1 & 3.0  & 8   & 1 & 0 & 0 & 3                      & 4                      & 4.5  & 2   & 0 & 0 & 1 & 1 & 0 & 3.5  & 20  & 1 & 1 & 7  & 6  & 5   \\
                                                                 & Diversity of materials/platforms      & 10  & 0 & 1 & 5 & 4 & 0 & 3.0  & 8   & 0 & 0 & 5 & 1                      & 2                      & 3.0  & 2   & 0 & 0 & 1 & 1 & 0 & 3.5  & 20  & 0 & 1 & 11 & 6  & 2   \\ 
\hline
\multirow{2}{*}{Objectives}                                      & Are the objectives met?               & 10  & 0 & 0 & 0 & 3 & 7 & 5.0  & 8   & 0 & 0 & 0 & 0                      & 8                      & 5.0  & 2   & 0 & 0 & 0 & 1 & 1 & 4.5  & 20  & 0 & 0 & 0  & 4  & 16  \\
                                                                 & Satisfaction rating                   & 10  & 0 & 0 & 0 & 2 & 8 & 5.0  & 8   & 0 & 0 & 0 & 2                      & 6                      & 5.0  & 2   & 0 & 0 & 1 & 0 & 1 & 4.0  & 20  & 0 & 0 & 1  & 4  & 15  \\
\hline
\end{tabular}
\end{sidewaystable*}

For commitment, most enrollees (9/10 enrollees) gave relatively high scores ($\geq$ 4) in their commitment in attending AC compared to volunteers (4/8 volunteers with scores $\geq$ 4). This can be due to AC being an elective class where mandatory attendance for enrollees boosts attendance rates \cite{Deenadayalan2008}. Our countercheck with Kruskal-Wallis test suggests that there is a significant difference in commitment in attending AC among enrollees, volunteers, and audiences, H(2) = 6.36, p = 0.042.  However, the pairwise post-hoc Dunn test with Bonferroni adjustments showed no significant difference in commitment in attending AC among each pair, despite enrollees and audiences showing a considerably low p-value (p = 0.089) (we set our significance level to 0.05). All learner types also show relatively low levels of participation during discussion (6/10 enrollees, 4/8 volunteers, and 2/2 audiences with scores $\leq$ 3). We see this as room for improvement in our AC in terms of participation among attendees and attending AC next semester. The former could imply that certain elements in our AC might have restricted discussion among participants (to be discussed in Sec.~\ref{sec:suggestions}). The latter might be a result of enrollees not needing the credits anymore next semester, and the volunteers and audiences being dominated by senior graduate students and professors, respectively. However, majority of our participants expressed their eagerness to join AC next semester (8/10 enrollees and 8/8 volunteers with scores $\geq$ 4).
The environment, on the other hand, received relatively high scores in almost all questions from all three groups ($\geq$ 6/10 enrollees and $\geq$ 7/8 volunteers with scores $\geq$ 4). This is verified by our counter check with Kruskal-Wallis and Mann-Whitney U tests, which suggest that there are no significant differences among the three groups in most of the questions under Environment. The absence of significance from our Kruskal-Wallis and Mann-Whitney U test results is likely due to our small sample size for each participant type, especially for the Audiences. However, volunteers gave a relatively low score (6/8 volunteers with scores $\leq$ 3) for their comfort in using VoxCharta compared to enrollees (2/10 with scores $\leq$ 3). Our countercheck with Mann-Whitney U test revealed that there is a significant difference between enrollees and volunteers in their comfort of using VoxCharta (U = 20, p = 0.035). This might indicate that those who are compelled to present papers as a requirement to pass the class find VoxCharta easier to navigate for reading new papers compared to volunteers who just present papers voluntarily. This may also reflect the volunteers' preference in using \texttt{arXiv} and other websites other than VoxCharta.

Furthermore, volunteers gave relatively high scores in the usefulness of gained knowledge (8/8 volunteers with scores $\geq$ 4), in comparison to enrollees (3/10 with scores $\geq$ 4), although the Kruskal-Wallis test results do not consider the difference to be significant (U = 4.63, p = 0.099) at 95\% significance level. The diversity of materials and platforms (e.g., websites, blogs, videos) discussed in our AC also needs improvement, which is evident from the relatively low scores from all participant types.

\begin{table*}
\caption{\label{tab:dunn}Pairwise post-hoc Dunn Test p-values for the questions that showed significant differences in the Kruskal-Wallis test in Table~\ref{tab:freq_dist}}
\begin{center}
\begin{ruledtabular}
\begin{tabular}{>{\centering\arraybackslash}m{8cm}>{\centering\arraybackslash}m{3cm}>{\centering\arraybackslash}m{3cm}>{\centering\arraybackslash}m{3cm}}
  \textbf{Pointers for each factor} & \textbf{Enrollees vs. Volunteers} & \textbf{Volunteers vs. Audiences} & \textbf{Enrollees vs. Audiences}  \\ \hline
  Commitment in attending AC & 0.207 & 0.895 & 0.089 \\ \cline{1-4}
  Comfort in papers outside their field	& \textbf{0.041} & 1.00 & 0.261
\end{tabular}
\begin{flushleft}
\textbf{Bold} indicates significant results (p $\leq$ 0.05) \\
\end{flushleft}
\end{ruledtabular}
\end{center}
\end{table*}

\begin{figure}[b]
\includegraphics[width=\columnwidth]{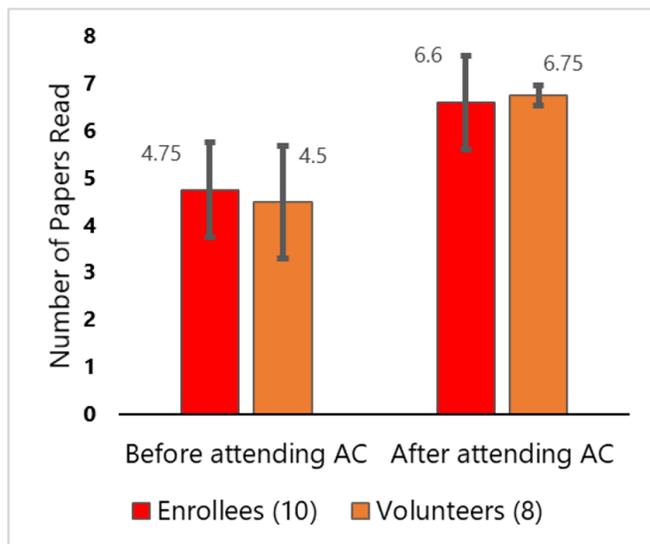}
\caption{Average number of papers read by enrollees (red) and volunteers (orange) before and after attending AC for the whole period of the Spring Semester 2020 (March 2020 - June 2020). Standard error bars are displayed.}
    \label{fig:average_number_of_papers} 
\end{figure}

\begin{figure*}
\includegraphics[width=2\columnwidth]{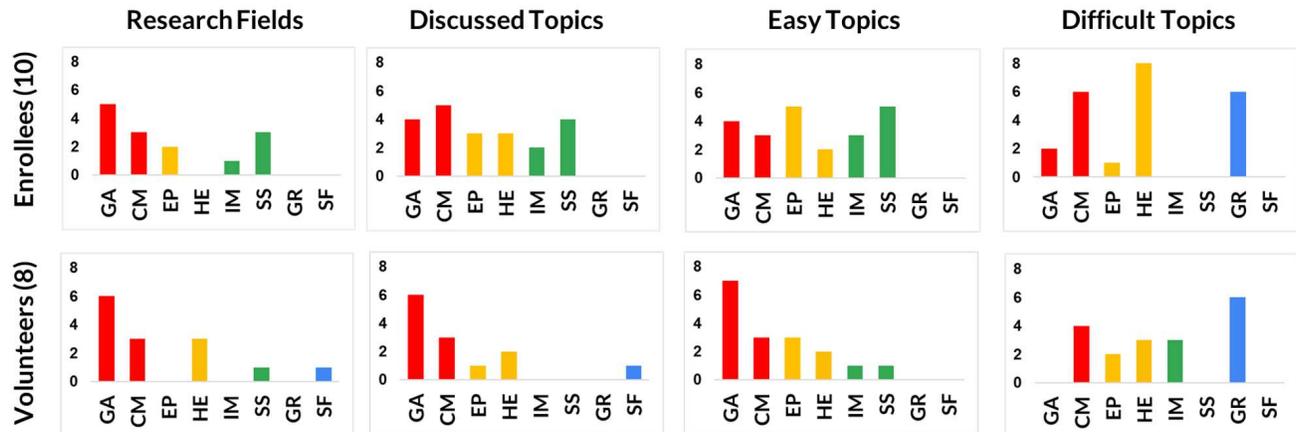}
\caption{\label{fig:topics} Bar chart showing the frequency distribution of research fields, discussed topics, topics that are easy, and topics that are difficult for enrollees (top) and volunteers (bottom). The number in the parenthesis refers to the sample size of each participant type. The research fields/topics are: Astrophysics of Galaxies (GA), Cosmology and Non-galactic Astronomy (CM), Earth and Planetary Astrophysics (EP), High Energy Astrophysical Phenomena (HE), Instrumentation and Methods for Astrophysics (IM), Solar and Stellar Astrophysics (SS), General Relativity and Quantum Cosmology (GR), Star Formation (SF). The research fields are based on arXiv's \textit{astro-ph} section, except for SF, which is answered collectively by 2 attendees in our AC. Topics with bars of the same color are added together to produce a 2$\times$4 matrix for a two-tailed Fisher exact test.}
\end{figure*}

As for the content of AC, we found that the number of papers read after joining AC increased for most attendees: 4.75 to 6.6 for enrollees and 4.5 to 6.75 for volunteers during the whole period of the Spring Semester 2020 (Fig.~\ref{fig:average_number_of_papers}). Enrollees gave relatively low scores for their comfort in reading papers outside their field (8/10 with scores $\leq$ 3) compared to volunteers (2/8 with scores $\leq$ 3). Our countercheck with Kruskal-Wallis test,  H(2) = 7.24, p = 0.027, and a pairwise post-hoc Dunn test with Bonferroni adjustments (p = 0.041) suggests that this may be significant. This is probably because most volunteers are senior graduate students and postdocs, so they may already have the prerequisite knowledge to understand papers outside of their fields, compared to enrollees who are mostly junior graduate students and undergraduates. We further investigated their comfort in reading papers outside their field in Fig.~\ref{fig:topics}. The Likert-scale questionnaire results also show relatively low scores from all groups (6/10 enrollees, 5/8 volunteers, and 1/2 audiences with scores $\leq$ 3) in the diversity of platforms used during discussion (i.e., videos, online news, slides, etc.), indicating this as another room for improvement in our AC.

Fig~\ref{fig:topics} shows a summary of the participants' research fields, discussed topics, easy topics, and difficult topics per participant type. We used a two-tailed Fisher test in a 2$\times$4 contingency table to evaluate whether there are any significant differences between their research fields and discussed topics, research fields and easy topics, and research fields and difficult topics. Since there are 8 research fields in total, we combined pairs of adjacent topics (bars with similar colors in Fig.~\ref{fig:topics}) to limit the number of topics from 8 to 4.
\begin{table*}
\caption{\label{tab:fisher}Two-tailed Fisher exact test p-values corresponding to the data in Fig.~\ref{fig:topics}.}
\begin{center}
\begin{ruledtabular}
\begin{tabular}{>{\centering\arraybackslash}m{8cm}>{\centering\arraybackslash}m{3cm}>{\centering\arraybackslash}m{3cm}>{\centering\arraybackslash}m{3cm}}
  \textbf{Compared groups} & \textbf{Enrollees} & \textbf{Volunteers} & \textbf{Audiences}  \\ \hline
  Research Field vs. Discussed Topics & 0.609 & 0.999 & 0.999 \\ \cline{1-4}
  Research Field vs. Easy Topics & 0.285 & 0.807 & 1.000 \\ \cline{1-4}
  Research Field vs. Difficult Topics & \textbf{0.003} & 0.095 & 1.000
\end{tabular}
\begin{flushleft}
\textbf{Bold} indicates significant results (p $\leq$ 0.05) \\
\end{flushleft}
\end{ruledtabular}
\end{center}
\end{table*}

Table~\ref{tab:fisher} shows the p-values from our two-tailed Fisher tests. 
It is clear that, for all participant types, their research fields and discussed topics have similar distributions, and so do their research fields and the topics that they find easy to understand. 
However, very small p-values arise when we compared the enrollees' and volunteers' distribution of research fields and topics that they find difficult to understand, indicating that these two distributions may be significantly different. 
Unfortunately, we were not able to derive statistically significant results for the audiences because of their very low sample size. Nevertheless, our results show that most participants tend to discuss papers in their research fields as they find topics outside their research field difficult to understand. As the research fields of our participants are diverse, this leads us to naturally discuss papers in many fields.

For topic bias, \cite{Bimczok2020} obtained a similar result with their inverted format of biology journal club. They required students to present papers related to their researches. This topic-centered approach is shown to motivate students and naturally broaden discussions due to the variety of methodological approaches presented in each paper. This also provided training for students to learn a variety of research fields which is advantageous in graduate-level courses. However, due to the diverse topics discussed, this approach limits in-depth discussions between individuals and more detailed parts of the papers. One particular difference of this work from ours is that we did not require the students to discuss papers about their research or research field. This implies that participants, by nature, are more inclined to discuss papers related to their research fields. As a result, much like what \cite{Bimczok2020} also showed, this enabled discussing a variety of topics while compromising in-depth discussions.

Finally, as for the objectives of our AC, all of our respondents greatly believe that our objectives are met. In addition, a high overall satisfaction rating for our AC was achieved on all learner types, indicating the success of our AC in achieving our goals.

\subsubsection{Suggestions and Comments from Participants}
\label{sec:suggestions}

We discuss the qualitative data that we gathered to support our quantitative data in the second phase of our study. 
Table~\ref{tab:comments} summarizes the key findings from our AC participants' comments. Our objectives were achieved based on their satisfactory comments. Furthermore, the diversity of topics in our AC can help people to learn other research fields. However, there is still room for improvement. For instance, due to the COVID-19 pandemic, we were not able to eat during AC, preventing us from assessing how food affects our AC. The lack of experts (e.g. postdoctoral researchers, faculty) also affected the quality of discussion in our AC. The lack of participation is also evident, which may be attributed to only allowing questions after the paper presentation. 
These findings were also apparent in our quantitative results in Table~\ref{tab:freq_dist}, where we achieved relatively low scores in participation during discussion. 
Lastly, because we used both VoxCharta and Facebook to post papers for discussion, some people find the two platforms difficult to navigate at the same time. This can support our significant statistical result between enrollees and volunteers in using VoxCharta in Table~\ref{tab:freq_dist}.

\begin{table*}
\caption{\label{tab:comments}Key findings from the comments and suggestions that we gathered from our participants. The factors where the key findings fall into and some direct quotes from the participants are also shown below. The number of suggestions/responses coded for each key finding are shown in the rightmost column.}
\begin{center}
\begin{ruledtabular}
\begin{tabular}{cccc}
  \textbf{Key Finding} & \textbf{Factor} & \textbf{Quotes} & \textbf{Number} \\ \hline
  Achieved objective & Objective & \makecell{``Personally I think the class helps \\ to summarise papers quickly..."}  & 1\\ \hline
  Satisfactory implementation & Objective & \makecell{``I think the class was good, and a \\ really important activity."} & 1\\ \hline
  Advantage of diverse topics & Content & \makecell{``I want to find a new exciting project \\ through this class. In this sense the diverse topics \\ are helpful to me."} & 3 \\ \hline
  Lack of food due to pandemic & Environment & \makecell{``Hope the coronavirus goes away soon so we can\\  eat in class..." \\ ``I like having lunch/Ubereats during astro-ph \\ to save time..."} & 2 \\ \hline
  Lack of experts & Commitment & \makecell{``If more postdoctoral can attend and maybe \\ give some suggestion for different field, that will \\ be great." \\ ``I think it would be even better if more \\ senior students/post-docs/staff were able to be \\ there too with a range of backgrounds and expertise, to \\ boost the discussion/experience and debate"} & 2\\ \hline
  Lack of participation & Commitment & \makecell{``Asking more questions during the paper presentation \\ rather than after each paper would allow for more \\ dynamic discussion"} & 1 \\ \hline
  Disadvantage of diverse platforms & Content & \makecell{``I don't think the current way of posting papers \\ to present is ideal. Can be hard to find at a later point, \\ and having two locations to post papers is sometimes \\ confusing if people post to one and not  the other."} & 2
\end{tabular}
\end{ruledtabular}
\end{center}
\end{table*}

\subsubsection{Limitations and Recommendations}
\label{sec:limitations}

Aside from food, there are other aspects of AC that we were not able to evaluate. One is the format of AC implementation. In our study, the AC being assessed is an elective class. Previous studies \citep[e.g.][]{Burris2019} and our qualitative phase results suggested that holding a journal club as a requirement (i.e. credit class) greatly improves the commitment and learning of participants, especially for the enrollees. Other formats of journal club implementation (e.g., as a non-credit course, credit course with a different grading system, online journal club, flipped classroom format, etc.) should be investigated in the future to widen perspectives on how to successfully hold an AC. 
Factors such as institutional culture and availability of resources, which we have not considered here, may also influence how AC is being implemented.

Note that the sample sizes for the two phases of our research method are low, especially for our triangulated phase. Although our AC is offered as a graduate elective course, there are also two post-doctoral researchers, two undergraduates, and two professors from various educational backgrounds joining the community. Since undergraduate students can attend graduate courses in our university, the involvement of the two undergraduates in this study inspires us to encourage more undergraduate students to participate in AC in the future. Investigating the factors that affect the success of AC can benefit greatly from a larger sample size. Thus we hope to be able to gather more statistically significant results from a larger sample size by expanding the scope of our participants into different types (e.g., mixed groups based on educational background, etc.) for future work.

Lastly, our study focused on pinpointing the different factors that should be considered to implement an AC effectively for learning the latest knowledge in astronomy. Although there are areas for improvement in the AC implementation of our university, it is beyond the scope of our paper to further examine which changes in our implementation will enhance the effectiveness of our AC implementation. A separate investigative study can be dedicated to inspect these areas for improvement.

\section{\label{sec:conclusion}Conclusion}
An investigative study on the factors that influence the success of preprint journal clubs in astronomy, more commonly known as Astro-ph/Astro-Coffee (AC), and on the best format of AC in our university (NTHU) was implemented. Survey dissemination and follow-up interviews with the respondents led us to conclude that there are 4 important factors for a successful AC: commitment (i.e. dedication of organizer/s and participants in maintaining AC), environment (i.e. the overall façade of the AC as a conducive forum for learning), content (i.e. topics and platforms for discussion), and objective (i.e. goals of the AC) (see Fig.~\ref{fig:factors}, Table~\ref{tab:freq_dist}, and Table~\ref{tab:dunn}). 
We also assessed the AC in our university (NTHU) by looking into the factors per participant type: enrollees, volunteers, and audiences. Our survey results (Table~\ref{tab:freq_dist}) show that most of the important aspects for each factors are perceived similarly by different learner types. However, some factors are perceived differently by different learner types: participants who regularly present papers tend to be more committed to attending AC compared to those who do not, and enrollees, who are composed of undergraduates and junior graduate students, feel less comfortable in reading papers outside their research field compared to volunteers, who are composed of postdocs and senior graduate students. Most participants are also biased towards discussing their fields of interest and perceived difficulty in discussing papers outside their fields of interest. Most participants feel that the objectives of our AC are met, as also shown by the increase in the number of papers being read after attending the class and the high satisfaction rating of the class. Lastly, areas for improvement, such as the lack of food, participation, attending experts, and disadvantage of using too many platforms for reading papers, were also highlighted.
Implementing AC can be made at the convenience of everyone, and so as a final remark, we would like to suggest other universities/institutes to evaluate their respective ACs with commitment, environment, content, and objective, as the main principles of their evaluation. Our main recommendations in starting and implementing a successful AC can be summarized into four points:
\begin{itemize}
    \item Specify what you would like to achieve in attending and implementing AC \textit{(objective)}
    \item Have a core of people (and an organizer, if possible) who are dedicated to attend AC every session and participate regularly \textit{(commitment)}
    \item Achieve balance in having an in-depth discussion of the papers presented and casual communication among participants, avoiding pressure and stress \textit{(environment)}
    \item For each paper discussed, introduce the basic concepts and show other visual aids (videos, pictures from the internet, etc.) if necessary/possible \textit{(content)}
\end{itemize}
As for other particular ways of presenting papers and organizing AC, we refer the readers to previous studies \cite[e.g.][]{Lee2005, Deenadayalan2008} which have also discussed ways of how to effectively run a journal club, although their discussion is in the medical context. 

\begin{acknowledgments}
We thank Dr. Su-Yen Chen for helpful discussions, and the anonymous referees for their constructive comments and suggestions. This work was supported by the Taiwan Ministry of Education (MOE) under Grant PMS1080022. AYLO and TH were supported by the Centre for Informatics and Computation in Astronomy (CICA) at National Tsing Hua University (NTHU) through a grant from the Ministry of Education of the Republic of China (Taiwan). AYLO’s  visit to NTHU was also supported by the Ministry of Science and Technology of the ROC (Taiwan) grant 105-2119-M-007-028-MY3, kindly hosted by Prof. Albert Kong.

\end{acknowledgments}

\appendix

\section{\label{appendixA}Online Survey to International Institutions}

We present here the online survey that we sent to different universities and institutions around the world. First, we ask for the basic information of our respondents (i.e., name, email, institution, position, research field). Then we ask questions pertaining to their AC. We have two sets of questions depending on whether they have AC in their institution or not.

For respondents with AC, we ask the following questions:
\begin{itemize}
    \item Composition of regular members who attend AC
    \item Who are invited to join AC?
    \item Astronomy fields discussed in AC
    \item Websites where you check new papers 
    \item Parts of the paper you focus on (e.g., methodology, results, abstract, conclusion, etc.)
    \item Schedule, duration, and frequency of AC
    \item Do you get any financial support for AC? If so, what do you spend it on?
    \item What kinds of foods and drinks do you have during AC?
    \item Do you think most participants will still come if there are no foods or drinks?
    \item Do you invite a speaker for your AC?
    \item What is the title of the organizer?
    \item How is AC announced?
    \item Is AC a required subject for students?
    \item How do you conduct AC?
    \item What are the usual benefits of AC to participants?
    \item What are the advantages and disadvantages of the way your AC is conducted?
    \item Any suggestions/experiences about AC that you would like to share?
    \item Are you willing to meet us via Skype, Zoom, or other video call platforms for a follow-up interview? How about email interview?
\end{itemize}

For respondents without AC, we ask the following questions:
\begin{itemize}
    \item Other events implemented in your institution related to astronomy (e.g., colloquium, seminars, symposium)
    \item Why AC is not conducted in your institution
    \item Have you ever thought about conducting AC in your institution (If yes, please describe some difficulties you encountered when you tried to host the activity. If no, please let us know the reason)
\end{itemize}

\section{\label{appendixB} Interview Protocol}

We present here the interview protocol we used for both follow-up video and email interviews. The interviews are semi-structured, meaning that new ideas might be brought up by the respondents during interview, allowing our interviews to be spontaneous and flexible. Here we show our frequently asked questions during the interview.

\begin{enumerate}
    \item Background of respondent 
    \begin{itemize}
        \item Name, affiliation, title
        \item Research field
        \item Do you organize AC in your institute?
    \end{itemize}
    \item Description of AC in your institute
    \begin{itemize}
        \item What has motivated you to organize AC? (for organizers)
        \item Who is/are the organizer(s) of your AC? (for non-organizers)
        \item What are your future plans for your AC?
        \item Are you satisfied with the way you are currently conducting AC? Why or why not?
        \item Do you think your current schedule of AC is effective? Why or why not?
        \item How long have you been doing AC in your institute?
        \item Are there more ACs in your institute?
        \item Why do you think that only a fraction of the people in your institute attend AC?
        \item Please briefly describe to us the environment on how you conduct your AC (e.g., casual roundtable discussion, classroom setup, food is served anytime, etc.)
        \item What are the difficulties you encountered as you first started your AC activities? Do you still encounter them from time to time?
        \item Do you agree that the position of the organizer (e.g., faculty, postdoc, or student) is important? Why or why not?
        \item Do you have any advice for those who would like to start an AC activity in their institutions?
        \item How many papers do you usually discuss per session? How long is each paper discussed?
        \item What do you think is the main purpose of your AC?
    \end{itemize}
    \item Follow-up Questions from the Online Survey Results
    \begin{itemize}
        \item Based on your survey response, you mentioned about your AC being a part of your curriculum (as an elective or required class). Could you explain further how this works?
        \item Based on your survey response, you mentioned about using emails in disseminating announcements about your AC. Could you tell us what your email announcements contain?
        \item Based on your survey response, you mentioned about your AC being limited to a certain demographic (e.g., graduate students only or faculty and postdocs only). Why is it so? 
        \item How are you able to invite participants from different research groups to attend you AC? And how do you handle the diverse research fields of your attendees during discussion?
        \item Based on your survey response, you mentioned about having food/drinks during AC. Who buys them? How do you pay for it? Have you tried not having them during AC? What is the effect?
        \item How do you discuss various fields in astronomy during AC?
    \end{itemize}
\end{enumerate}


\nocite{*}

\input{output.bbl}
\end{document}

%% file: output.bbl
\providecommand{\noopsort}[1]{}\providecommand{\singleletter}[1]{#1}%

%% file: APS_MainText.bbl
\begin{thebibliography}{61}%
\makeatletter
\providecommand \@ifxundefined [1]{%
 \@ifx{#1\undefined}
}%
\providecommand \@ifnum [1]{%
 \ifnum #1\expandafter \@firstoftwo
 \else \expandafter \@secondoftwo
 \fi
}%
\providecommand \@ifx [1]{%
 \ifx #1\expandafter \@firstoftwo
 \else \expandafter \@secondoftwo
 \fi
}%
\providecommand \natexlab [1]{#1}%
\providecommand \enquote  [1]{``#1''}%
\providecommand \bibnamefont  [1]{#1}%
\providecommand \bibfnamefont [1]{#1}%
\providecommand \citenamefont [1]{#1}%
\providecommand \href@noop [0]{\@secondoftwo}%
\providecommand \href [0]{\begingroup \@sanitize@url \@href}%
\providecommand \@href[1]{\@@startlink{#1}\@@href}%
\providecommand \@@href[1]{\endgroup#1\@@endlink}%
\providecommand \@sanitize@url [0]{\catcode `\\12\catcode `\$12\catcode
  `\&12\catcode `\#12\catcode `\^12\catcode `\_12\catcode `\%12\relax}%
\providecommand \@@startlink[1]{}%
\providecommand \@@endlink[0]{}%
\providecommand \url  [0]{\begingroup\@sanitize@url \@url }%
\providecommand \@url [1]{\endgroup\@href {#1}{\urlprefix }}%
\providecommand \urlprefix  [0]{URL }%
\providecommand \Eprint [0]{\href }%
\providecommand \doibase [0]{https://doi.org/}%
\providecommand \selectlanguage [0]{\@gobble}%
\providecommand \bibinfo  [0]{\@secondoftwo}%
\providecommand \bibfield  [0]{\@secondoftwo}%
\providecommand \translation [1]{[#1]}%
\providecommand \BibitemOpen [0]{}%
\providecommand \bibitemStop [0]{}%
\providecommand \bibitemNoStop [0]{.\EOS\space}%
\providecommand \EOS [0]{\spacefactor3000\relax}%
\providecommand \BibitemShut  [1]{\csname bibitem#1\endcsname}%
\let\auto@bib@innerbib\@empty
\bibitem [{\citenamefont {Einstein}\ and\ \citenamefont
  {Rosen}(1937)}]{Einstein1937}%
  \BibitemOpen
  \bibfield  {author} {\bibinfo {author} {\bibfnamefont {A.}~\bibnamefont
  {Einstein}}\ and\ \bibinfo {author} {\bibfnamefont {N.}~\bibnamefont
  {Rosen}},\ }\bibfield  {title} {\bibinfo {title} {On gravitational waves},\
  }\href@noop {} {\bibfield  {journal} {\bibinfo  {journal} {J. Franklin
  Inst.}\ }\textbf {\bibinfo {volume} {223}},\ \bibinfo {pages} {43} (\bibinfo
  {year} {1937})}\BibitemShut {NoStop}%
\bibitem [{\citenamefont {Abbott}(2016)}]{Abbott2016}%
  \BibitemOpen
  \bibfield  {author} {\bibinfo {author} {\bibfnamefont {B.~P.}\ \bibnamefont
  {Abbott}} (\bibinfo {collaboration} {LIGO Scientific Collaboration and Virgo
  Collaboration}),\ }\bibfield  {title} {\bibinfo {title} {Observation of
  gravitational waves from a binary black hole merger},\ }\href@noop {}
  {\bibfield  {journal} {\bibinfo  {journal} {Phys. \ Rev. \ Lett.}\ }\textbf
  {\bibinfo {volume} {116}} (\bibinfo {year} {2016})}\BibitemShut {NoStop}%
\bibitem [{\citenamefont {Granot}\ \emph {et~al.}(2017)\citenamefont {Granot},
  \citenamefont {Guetta},\ and\ \citenamefont {Gill}}]{Granot2017}%
  \BibitemOpen
  \bibfield  {author} {\bibinfo {author} {\bibfnamefont {J.}~\bibnamefont
  {Granot}}, \bibinfo {author} {\bibfnamefont {D.}~\bibnamefont {Guetta}},\
  and\ \bibinfo {author} {\bibfnamefont {R.}~\bibnamefont {Gill}},\ }\bibfield
  {title} {\bibinfo {title} {Lessons from the short {GRB} 170817{A} – the
  first gravitational wave detection of a binary neutron star merger},\
  }\href@noop {} {\bibfield  {journal} {\bibinfo  {journal} {ApJ}\ }\textbf
  {\bibinfo {volume} {850}},\ \bibinfo {pages} {L24} (\bibinfo {year}
  {2017})}\BibitemShut {NoStop}%
\bibitem [{\citenamefont {Ade}(014a)}]{Ade2014a}%
  \BibitemOpen
  \bibfield  {author} {\bibinfo {author} {\bibfnamefont {P.~A.~R.}\
  \bibnamefont {Ade}} (\bibinfo {collaboration} {Planck Collaboration}),\
  }\bibfield  {title} {\bibinfo {title} {Planck 2013 results. {XVI}.
  {C}osmological parameters},\ }\href@noop {} {\bibfield  {journal} {\bibinfo
  {journal} {A. \& A.}\ }\textbf {\bibinfo {volume} {571}},\ \bibinfo {pages}
  {A16} (\bibinfo {year} {2014a})}\BibitemShut {NoStop}%
\bibitem [{\citenamefont {Ade}(014b)}]{Ade2014b}%
  \BibitemOpen
  \bibfield  {author} {\bibinfo {author} {\bibfnamefont {P.~A.~R.}\
  \bibnamefont {Ade}} (\bibinfo {collaboration} {POLARBEAR Collaboration}),\
  }\bibfield  {title} {\bibinfo {title} {A measurement of the cosmic microwave
  background {B}-mode polarization power spectrum at sub-degree scales with
  {POLARBEAR}},\ }\href@noop {} {\bibfield  {journal} {\bibinfo  {journal}
  {ApJ}\ }\textbf {\bibinfo {volume} {794}},\ \bibinfo {pages} {171} (\bibinfo
  {year} {2014b})}\BibitemShut {NoStop}%
\bibitem [{\citenamefont {Clampin}(2008)}]{Clampin2008}%
  \BibitemOpen
  \bibfield  {author} {\bibinfo {author} {\bibfnamefont {M.}~\bibnamefont
  {Clampin}},\ }\bibfield  {title} {\bibinfo {title} {The {J}ames {W}ebb
  {S}pace {T}elescope ({JWST})},\ }\href@noop {} {\bibfield  {journal}
  {\bibinfo  {journal} {ASR}\ }\textbf {\bibinfo {volume} {41}},\ \bibinfo
  {pages} {1983} (\bibinfo {year} {2008})}\BibitemShut {NoStop}%
\bibitem [{\citenamefont {Gilmozzi}\ and\ \citenamefont
  {Spyromilio}(2007)}]{Gilmozzi2007}%
  \BibitemOpen
  \bibfield  {author} {\bibinfo {author} {\bibfnamefont {R.}~\bibnamefont
  {Gilmozzi}}\ and\ \bibinfo {author} {\bibfnamefont {J.}~\bibnamefont
  {Spyromilio}},\ }\bibfield  {title} {\bibinfo {title} {The {E}uropean
  {E}xtremely {L}arge {T}elescope ({E-ELT})},\ }\href@noop {} {\bibfield
  {journal} {\bibinfo  {journal} {The Messenger}\ }\textbf {\bibinfo {volume}
  {127}},\ \bibinfo {pages} {3} (\bibinfo {year} {2007})}\BibitemShut {NoStop}%
\bibitem [{\citenamefont {Sanders}(2012)}]{Sanders2013}%
  \BibitemOpen
  \bibfield  {author} {\bibinfo {author} {\bibfnamefont {G.~H.}\ \bibnamefont
  {Sanders}},\ }\bibfield  {title} {\bibinfo {title} {The {Thirty Meter
  Telescope (TMT)}: An international observatory},\ }\href@noop {} {\bibfield
  {journal} {\bibinfo  {journal} {J. Astrophys. Astron.}\ }\textbf {\bibinfo
  {volume} {34}},\ \bibinfo {pages} {81} (\bibinfo {year} {2012})}\BibitemShut
  {NoStop}%
\bibitem [{\citenamefont {Dwarakanath}\ and\ \citenamefont
  {Khan}(2000)}]{Dwarakanath2000}%
  \BibitemOpen
  \bibfield  {author} {\bibinfo {author} {\bibfnamefont {L.~S.}\ \bibnamefont
  {Dwarakanath}}\ and\ \bibinfo {author} {\bibfnamefont {K.~S.}\ \bibnamefont
  {Khan}},\ }\bibfield  {title} {\bibinfo {title} {Modernizing the journal
  club},\ }\href@noop {} {\bibfield  {journal} {\bibinfo  {journal} {Br. J.
  Hosp. Med.}\ }\textbf {\bibinfo {volume} {61}},\ \bibinfo {pages} {425}
  (\bibinfo {year} {2000})}\BibitemShut {NoStop}%
\bibitem [{\citenamefont {Topf}\ \emph {et~al.}(2017)\citenamefont {Topf},
  \citenamefont {Sparks}, \citenamefont {Phelan}, \citenamefont {Shah},
  \citenamefont {Lerma}, \citenamefont {Graham-Brown}, \citenamefont
  {Madariaga}, \citenamefont {Iannuzzella}, \citenamefont {Rheault},
  \citenamefont {Oates},\ and\ \citenamefont {et~al.}}]{Topf2017}%
  \BibitemOpen
  \bibfield  {author} {\bibinfo {author} {\bibfnamefont {J.~M.}\ \bibnamefont
  {Topf}}, \bibinfo {author} {\bibfnamefont {M.~A.}\ \bibnamefont {Sparks}},
  \bibinfo {author} {\bibfnamefont {P.~J.}\ \bibnamefont {Phelan}}, \bibinfo
  {author} {\bibfnamefont {N.}~\bibnamefont {Shah}}, \bibinfo {author}
  {\bibfnamefont {E.~V.}\ \bibnamefont {Lerma}}, \bibinfo {author}
  {\bibfnamefont {M.~P.~M.}\ \bibnamefont {Graham-Brown}}, \bibinfo {author}
  {\bibfnamefont {H.}~\bibnamefont {Madariaga}}, \bibinfo {author}
  {\bibfnamefont {F.}~\bibnamefont {Iannuzzella}}, \bibinfo {author}
  {\bibfnamefont {M.~N.}\ \bibnamefont {Rheault}}, \bibinfo {author}
  {\bibfnamefont {T.}~\bibnamefont {Oates}},\ and\ \bibinfo {author}
  {\bibnamefont {et~al.}},\ }\bibfield  {title} {\bibinfo {title} {The
  evolution of the journal club: {F}rom osler to twitter},\ }\href@noop {}
  {\bibfield  {journal} {\bibinfo  {journal} {Am. J. Kidney Dis.}\ }\textbf
  {\bibinfo {volume} {69}},\ \bibinfo {pages} {827} (\bibinfo {year}
  {2017})}\BibitemShut {NoStop}%
\bibitem [{\citenamefont {Linzer}(1987)}]{Linzer1987}%
  \BibitemOpen
  \bibfield  {author} {\bibinfo {author} {\bibfnamefont {M.}~\bibnamefont
  {Linzer}},\ }\bibfield  {title} {\bibinfo {title} {The journal club and
  medical education: {O}ver one hundred years of unrecorded history},\
  }\href@noop {} {\bibfield  {journal} {\bibinfo  {journal} {Postgrad. Med.
  J.}\ }\textbf {\bibinfo {volume} {63}},\ \bibinfo {pages} {475} (\bibinfo
  {year} {1987})}\BibitemShut {NoStop}%
\bibitem [{\citenamefont {Mazal}\ and\ \citenamefont
  {Truluck}(2014)}]{Mazal2014}%
  \BibitemOpen
  \bibfield  {author} {\bibinfo {author} {\bibfnamefont {J.}~\bibnamefont
  {Mazal}}\ and\ \bibinfo {author} {\bibfnamefont {C.}~\bibnamefont
  {Truluck}},\ }\bibfield  {title} {\bibinfo {title} {Organizing and leading a
  journal club},\ }\href@noop {} {\bibfield  {journal} {\bibinfo  {journal}
  {Radiol. Technol}\ }\textbf {\bibinfo {volume} {85}},\ \bibinfo {pages} {549}
  (\bibinfo {year} {2014})}\BibitemShut {NoStop}%
\bibitem [{\citenamefont {Brown}(999a)}]{Brown1999a}%
  \BibitemOpen
  \bibfield  {author} {\bibinfo {author} {\bibfnamefont {C.~M.}\ \bibnamefont
  {Brown}},\ }\bibfield  {title} {\bibinfo {title} {Information literacy of
  physical science graduate students in the information age},\ }\href@noop {}
  {\bibfield  {journal} {\bibinfo  {journal} {Coll. Res. Libr.}\ }\textbf
  {\bibinfo {volume} {60}},\ \bibinfo {pages} {426} (\bibinfo {year}
  {1999a})}\BibitemShut {NoStop}%
\bibitem [{\citenamefont {Jamali}\ and\ \citenamefont
  {Nicholas}(2008)}]{Jamali2008}%
  \BibitemOpen
  \bibfield  {author} {\bibinfo {author} {\bibfnamefont {H.~R.}\ \bibnamefont
  {Jamali}}\ and\ \bibinfo {author} {\bibfnamefont {D.}~\bibnamefont
  {Nicholas}},\ }\bibfield  {title} {\bibinfo {title} {Communication and
  information‐seeking behavior of {PhD} students in physicists and
  astronomy},\ }\href@noop {} {\bibfield  {journal} {\bibinfo  {journal}
  {Proceedings of the American Society for Information Science and Technology}\
  }\textbf {\bibinfo {volume} {43}},\ \bibinfo {pages} {1} (\bibinfo {year}
  {2008})}\BibitemShut {NoStop}%
\bibitem [{\citenamefont {Cho}(2020)}]{Cho2000}%
  \BibitemOpen
  \bibfield  {author} {\bibinfo {author} {\bibfnamefont {A.}~\bibnamefont
  {Cho}},\ }\bibfield  {title} {\bibinfo {title} {Distorted galaxies point to
  dark matter},\ }\href@noop {} {\bibfield  {journal} {\bibinfo  {journal}
  {Science}\ }\textbf {\bibinfo {volume} {287}},\ \bibinfo {pages} {1899}
  (\bibinfo {year} {2020})}\BibitemShut {NoStop}%
\bibitem [{\citenamefont {Kling}\ and\ \citenamefont
  {McKim}(2000)}]{Kling2000}%
  \BibitemOpen
  \bibfield  {author} {\bibinfo {author} {\bibfnamefont {R.}~\bibnamefont
  {Kling}}\ and\ \bibinfo {author} {\bibfnamefont {G.}~\bibnamefont {McKim}},\
  }\bibfield  {title} {\bibinfo {title} {Not just a matter of time: field
  differences in the shaping of electronic media in supporting scientific
  communication},\ }\href@noop {} {\bibfield  {journal} {\bibinfo  {journal}
  {J. Assoc. Inf. Sci. Technol.}\ }\textbf {\bibinfo {volume} {51}},\ \bibinfo
  {pages} {1306} (\bibinfo {year} {2000})}\BibitemShut {NoStop}%
\bibitem [{\citenamefont {Fry}(2003)}]{Fry2003}%
  \BibitemOpen
  \bibfield  {author} {\bibinfo {author} {\bibfnamefont {J.}~\bibnamefont
  {Fry}},\ }\emph {\bibinfo {title} {Cultural shaping of scholarly
  communication within academic specialisms}},\ \href@noop {} {\bibinfo {type}
  {{Ph.D.} thesis}},\ \bibinfo  {school} {University of Brighton} (\bibinfo
  {year} {2003})\BibitemShut {NoStop}%
\bibitem [{\citenamefont {Lawal}(2002)}]{Lawal2002}%
  \BibitemOpen
  \bibfield  {author} {\bibinfo {author} {\bibfnamefont {I.}~\bibnamefont
  {Lawal}},\ }\bibfield  {title} {\bibinfo {title} {Scholarly communication:
  {T}he use and non-use of e-print archives for the dissemination of scientific
  information},\ }\href@noop {} {\bibfield  {journal} {\bibinfo  {journal}
  {Issues Sci. Technol. Librariansh.}\ } (\bibinfo {year} {2002})},\ \bibinfo
  {note} {dOI:10.5062/F4057CWP}\BibitemShut {NoStop}%
\bibitem [{\citenamefont {Lim}(1996)}]{Lim1996}%
  \BibitemOpen
  \bibfield  {author} {\bibinfo {author} {\bibfnamefont {E.}~\bibnamefont
  {Lim}},\ }\bibfield  {title} {\bibinfo {title} {Preprint servers: {A} new
  model for scholarly publishing?},\ }\href@noop {} {\bibfield  {journal}
  {\bibinfo  {journal} {Aust. Acad. Res. Libr.}\ }\textbf {\bibinfo {volume}
  {27}},\ \bibinfo {pages} {21} (\bibinfo {year} {1996})}\BibitemShut {NoStop}%
\bibitem [{\citenamefont {Brown}(001a)}]{Brown2001a}%
  \BibitemOpen
  \bibfield  {author} {\bibinfo {author} {\bibfnamefont {C.~M.}\ \bibnamefont
  {Brown}},\ }\bibfield  {title} {\bibinfo {title} {The coming of age of
  e-prints in the literature of physics},\ }\href@noop {} {\bibfield  {journal}
  {\bibinfo  {journal} {Issues Sci. Technol. Librariansh.}\ } (\bibinfo {year}
  {2001a})}\BibitemShut {NoStop}%
\bibitem [{\citenamefont {Brown}(001b)}]{Brown2001b}%
  \BibitemOpen
  \bibfield  {author} {\bibinfo {author} {\bibfnamefont {C.~M.}\ \bibnamefont
  {Brown}},\ }\bibfield  {title} {\bibinfo {title} {The e‐volution of
  preprints in the scholarly communication of physicists and astronomer},\
  }\href@noop {} {\bibfield  {journal} {\bibinfo  {journal} {J. Assoc. Inf.
  Sci. Technol.}\ }\textbf {\bibinfo {volume} {52}},\ \bibinfo {pages} {187}
  (\bibinfo {year} {2001b})}\BibitemShut {NoStop}%
\bibitem [{\citenamefont {Casadevall}\ and\ \citenamefont
  {Gow}(2018)}]{Casadevall2018}%
  \BibitemOpen
  \bibfield  {author} {\bibinfo {author} {\bibfnamefont {A.}~\bibnamefont
  {Casadevall}}\ and\ \bibinfo {author} {\bibfnamefont {N.}~\bibnamefont
  {Gow}},\ }\bibfield  {title} {\bibinfo {title} {Using preprints for journal
  clubs},\ }\href@noop {} {\bibfield  {journal} {\bibinfo  {journal} {Am. \
  Soc. \ Microbiol.}\ }\textbf {\bibinfo {volume} {9}} (\bibinfo {year}
  {2018})},\ \bibinfo {note} {{DOI:10.1128/mBio.00516-18}}\BibitemShut
  {NoStop}%
\bibitem [{\citenamefont {Avasthi}\ \emph {et~al.}(2018)\citenamefont
  {Avasthi}, \citenamefont {Soragni},\ and\ \citenamefont
  {Bembenek}}]{Avasthi2018}%
  \BibitemOpen
  \bibfield  {author} {\bibinfo {author} {\bibfnamefont {P.}~\bibnamefont
  {Avasthi}}, \bibinfo {author} {\bibfnamefont {A.}~\bibnamefont {Soragni}},\
  and\ \bibinfo {author} {\bibfnamefont {J.~N.}\ \bibnamefont {Bembenek}},\
  }\bibfield  {title} {\bibinfo {title} {Point of view: {J}ournal clubs in the
  time of preprints},\ }\href@noop {} {\bibfield  {journal} {\bibinfo
  {journal} {eLife}\ }\textbf {\bibinfo {volume} {7}},\ \bibinfo {pages}
  {e38532} (\bibinfo {year} {2018})}\BibitemShut {NoStop}%
\bibitem [{\citenamefont {Marra}(2018)}]{Marra2018}%
  \BibitemOpen
  \bibfield  {author} {\bibinfo {author} {\bibfnamefont {M.}~\bibnamefont
  {Marra}},\ }\bibfield  {title} {\bibinfo {title} {Astrophysicists and
  physicists as creators of {ArXiv}-based commenting resources for their
  research communities. {A}n initial survey},\ }\href@noop {} {\bibfield
  {journal} {\bibinfo  {journal} {Inf. Serv. Use}\ }\textbf {\bibinfo {volume}
  {37}},\ \bibinfo {pages} {371} (\bibinfo {year} {2018})}\BibitemShut
  {NoStop}%
\bibitem [{\citenamefont {.Wenger}(998a)}]{Wenger1998a}%
  \BibitemOpen
  \bibfield  {author} {\bibinfo {author} {\bibfnamefont {E.}~\bibnamefont
  {.Wenger}},\ }\href@noop {} {\emph {\bibinfo {title} {Communities of
  practice: Learning, meaning, and identity}}}\ (\bibinfo  {publisher} {New
  York: Cambridge University},\ \bibinfo {year} {1998a})\BibitemShut {NoStop}%
\bibitem [{\citenamefont {Newswander}\ and\ \citenamefont
  {Borrego}(2009)}]{Newswander2009}%
  \BibitemOpen
  \bibfield  {author} {\bibinfo {author} {\bibfnamefont {L.~K.}\ \bibnamefont
  {Newswander}}\ and\ \bibinfo {author} {\bibfnamefont {M.}~\bibnamefont
  {Borrego}},\ }\bibfield  {title} {\bibinfo {title} {Using journal clubs to
  cultivate a community of practice at the graduate level},\ }\href@noop {}
  {\bibfield  {journal} {\bibinfo  {journal} {Eur. J. Eng. Educ.}\ }\textbf
  {\bibinfo {volume} {34}},\ \bibinfo {pages} {561} (\bibinfo {year}
  {2009})}\BibitemShut {NoStop}%
\bibitem [{\citenamefont {Quinn}\ \emph {et~al.}(2014)\citenamefont {Quinn},
  \citenamefont {Cantillon}, \citenamefont {Redmond},\ and\ \citenamefont
  {Bennett}}]{Quinn2014}%
  \BibitemOpen
  \bibfield  {author} {\bibinfo {author} {\bibfnamefont {E.~M.}\ \bibnamefont
  {Quinn}}, \bibinfo {author} {\bibfnamefont {P.}~\bibnamefont {Cantillon}},
  \bibinfo {author} {\bibfnamefont {H.~P.}\ \bibnamefont {Redmond}},\ and\
  \bibinfo {author} {\bibfnamefont {D.}~\bibnamefont {Bennett}},\ }\bibfield
  {title} {\bibinfo {title} {Surgical journal club as a community of practice:
  a case study},\ }\href@noop {} {\bibfield  {journal} {\bibinfo  {journal} {J.
  Surg. Educ.}\ }\textbf {\bibinfo {volume} {71}},\ \bibinfo {pages} {606}
  (\bibinfo {year} {2014})}\BibitemShut {NoStop}%
\bibitem [{\citenamefont {Stein}(1998)}]{Stein1998}%
  \BibitemOpen
  \bibfield  {author} {\bibinfo {author} {\bibfnamefont {D.}~\bibnamefont
  {Stein}},\ }\href@noop {} {\emph {\bibinfo {title} {Situated learning in
  adult education}}}\ (\bibinfo  {publisher} {Ohio State University},\ \bibinfo
  {year} {1998})\BibitemShut {NoStop}%
\bibitem [{\citenamefont {Lave}\ and\ \citenamefont {Wenger}(1991)}]{Lave1991}%
  \BibitemOpen
  \bibfield  {author} {\bibinfo {author} {\bibfnamefont {J.}~\bibnamefont
  {Lave}}\ and\ \bibinfo {author} {\bibfnamefont {E.}~\bibnamefont {Wenger}},\
  }\href@noop {} {\emph {\bibinfo {title} {Situated learning: Legitimate
  peripheral participation}}}\ (\bibinfo  {publisher} {Cambridge University
  Press},\ \bibinfo {year} {1991})\BibitemShut {NoStop}%
\bibitem [{\citenamefont {Wenger}(998b)}]{Wenger1998b}%
  \BibitemOpen
  \bibfield  {author} {\bibinfo {author} {\bibfnamefont {E.}~\bibnamefont
  {Wenger}},\ }\bibfield  {title} {\bibinfo {title} {Communities of practice:
  Learning as a social system},\ }\href@noop {} {\bibfield  {journal} {\bibinfo
   {journal} {Syst Thinker}\ }\textbf {\bibinfo {volume} {9}},\ \bibinfo
  {pages} {2} (\bibinfo {year} {1998b})}\BibitemShut {NoStop}%
\bibitem [{\citenamefont {Chan}\ \emph {et~al.}(2015)\citenamefont {Chan},
  \citenamefont {Thoma}, \citenamefont {Radecki}, \citenamefont {Topf},
  \citenamefont {Woo}, \citenamefont {Kao}, \citenamefont {Cochran},
  \citenamefont {Hiremath},\ and\ \citenamefont {Lin}}]{Chan2015}%
  \BibitemOpen
  \bibfield  {author} {\bibinfo {author} {\bibfnamefont {T.~M.}\ \bibnamefont
  {Chan}}, \bibinfo {author} {\bibfnamefont {B.}~\bibnamefont {Thoma}},
  \bibinfo {author} {\bibfnamefont {R.}~\bibnamefont {Radecki}}, \bibinfo
  {author} {\bibfnamefont {J.}~\bibnamefont {Topf}}, \bibinfo {author}
  {\bibfnamefont {H.~H.}\ \bibnamefont {Woo}}, \bibinfo {author} {\bibfnamefont
  {L.~S.}\ \bibnamefont {Kao}}, \bibinfo {author} {\bibfnamefont
  {A.}~\bibnamefont {Cochran}}, \bibinfo {author} {\bibfnamefont
  {S.}~\bibnamefont {Hiremath}},\ and\ \bibinfo {author} {\bibfnamefont
  {M.}~\bibnamefont {Lin}},\ }\bibfield  {title} {\bibinfo {title} {Ten steps
  for setting up an online journal club},\ }\href@noop {} {\bibfield  {journal}
  {\bibinfo  {journal} {J. Contin. Educ. Health Prof}\ }\textbf {\bibinfo
  {volume} {35}},\ \bibinfo {pages} {148} (\bibinfo {year} {2015})}\BibitemShut
  {NoStop}%
\bibitem [{\citenamefont {Tallman}\ and\ \citenamefont
  {Feldman}(2016)}]{Tallman2016}%
  \BibitemOpen
  \bibfield  {author} {\bibinfo {author} {\bibfnamefont {K.~A.}\ \bibnamefont
  {Tallman}}\ and\ \bibinfo {author} {\bibfnamefont {A.}~\bibnamefont
  {Feldman}},\ }\bibfield  {title} {\bibinfo {title} {The use of journal clubs
  in science teacher education},\ }\href@noop {} {\bibfield  {journal}
  {\bibinfo  {journal} {JSTE}\ }\textbf {\bibinfo {volume} {27}},\ \bibinfo
  {pages} {325} (\bibinfo {year} {2016})}\BibitemShut {NoStop}%
\bibitem [{\citenamefont {Lin}\ and\ \citenamefont {Sherbino}(2015)}]{Lin2015}%
  \BibitemOpen
  \bibfield  {author} {\bibinfo {author} {\bibfnamefont {M.}~\bibnamefont
  {Lin}}\ and\ \bibinfo {author} {\bibfnamefont {J.}~\bibnamefont {Sherbino}},\
  }\bibfield  {title} {\bibinfo {title} {Creating a virtual journal club: A
  community of practice using multiple social media strategies},\ }\href@noop
  {} {\bibfield  {journal} {\bibinfo  {journal} {J. Grad. Med. Educ.}\ }\textbf
  {\bibinfo {volume} {7}},\ \bibinfo {pages} {481} (\bibinfo {year}
  {2015})}\BibitemShut {NoStop}%
\bibitem [{\citenamefont {Palicsar}(1998)}]{Palincsar1998}%
  \BibitemOpen
  \bibfield  {author} {\bibinfo {author} {\bibfnamefont {A.~S.}\ \bibnamefont
  {Palicsar}},\ }\bibfield  {title} {\bibinfo {title} {Social constructivist
  perspectives on teaching and learning},\ }\href@noop {} {\bibfield  {journal}
  {\bibinfo  {journal} {Annu. Rev. Psychol.}\ }\textbf {\bibinfo {volume}
  {49}},\ \bibinfo {pages} {345} (\bibinfo {year} {1998})}\BibitemShut
  {NoStop}%
\bibitem [{\citenamefont {Nesbitt}\ and\ \citenamefont
  {Barton}(2014)}]{Nesbitt2014}%
  \BibitemOpen
  \bibfield  {author} {\bibinfo {author} {\bibfnamefont {J.}~\bibnamefont
  {Nesbitt}}\ and\ \bibinfo {author} {\bibfnamefont {G.}~\bibnamefont
  {Barton}},\ }\bibfield  {title} {\bibinfo {title} {Nursing journal clubs: A
  strategy for improving knowledge translation and evidenced-informed clinical
  practice invited manuscript for the journal of radiology nursing},\
  }\href@noop {} {\bibfield  {journal} {\bibinfo  {journal} {J Radiol Nurs}\
  }\textbf {\bibinfo {volume} {33}},\ \bibinfo {pages} {3} (\bibinfo {year}
  {2014})}\BibitemShut {NoStop}%
\bibitem [{\citenamefont {Lachance}(2014)}]{Lachance2014}%
  \BibitemOpen
  \bibfield  {author} {\bibinfo {author} {\bibfnamefont {C.}~\bibnamefont
  {Lachance}},\ }\bibfield  {title} {\bibinfo {title} {Nursing journal clubs: A
  literature review on the effective teaching strategy for continuing education
  and evidence-based practice},\ }\href@noop {} {\bibfield  {journal} {\bibinfo
   {journal} {J Contin Educ Nurs}\ }\textbf {\bibinfo {volume} {45}},\ \bibinfo
  {pages} {559} (\bibinfo {year} {2014})}\BibitemShut {NoStop}%
\bibitem [{\citenamefont {Burls}(2014)}]{Burls2014}%
  \BibitemOpen
  \bibfield  {author} {\bibinfo {author} {\bibfnamefont {A.}~\bibnamefont
  {Burls}},\ }\href@noop {} {\emph {\bibinfo {title} {What is critical
  appraisal?}}}\ (\bibinfo  {publisher} {Hayward Medical Communications},\
  \bibinfo {year} {2014})\BibitemShut {NoStop}%
\bibitem [{\citenamefont {Khan}\ and\ \citenamefont {Gee}(1999)}]{Khan1999}%
  \BibitemOpen
  \bibfield  {author} {\bibinfo {author} {\bibfnamefont {K.~S.}\ \bibnamefont
  {Khan}}\ and\ \bibinfo {author} {\bibfnamefont {H.}~\bibnamefont {Gee}},\
  }\bibfield  {title} {\bibinfo {title} {A new approach to teaching and
  learning in journal club},\ }\href@noop {} {\bibfield  {journal} {\bibinfo
  {journal} {Med. Teach.}\ }\textbf {\bibinfo {volume} {21}},\ \bibinfo {pages}
  {289} (\bibinfo {year} {1999})}\BibitemShut {NoStop}%
\bibitem [{\citenamefont {Rosewall}(2012)}]{Rosewall2012}%
  \BibitemOpen
  \bibfield  {author} {\bibinfo {author} {\bibfnamefont {T.}~\bibnamefont
  {Rosewall}},\ }\bibfield  {title} {\bibinfo {title} {The use of journal clubs
  in canadian radiation therapy departments: {P}revalence and perceptions},\
  }\href@noop {} {\bibfield  {journal} {\bibinfo  {journal} {J. Med. Imaging
  Radiat. Sci.}\ }\textbf {\bibinfo {volume} {43}},\ \bibinfo {pages} {16}
  (\bibinfo {year} {2012})}\BibitemShut {NoStop}%
\bibitem [{\citenamefont {Milinkovic}\ \emph {et~al.}(2008)\citenamefont
  {Milinkovic}, \citenamefont {Field},\ and\ \citenamefont
  {Agustin}}]{Milinkovic2008}%
  \BibitemOpen
  \bibfield  {author} {\bibinfo {author} {\bibfnamefont {D.}~\bibnamefont
  {Milinkovic}}, \bibinfo {author} {\bibfnamefont {N.}~\bibnamefont {Field}},\
  and\ \bibinfo {author} {\bibfnamefont {C.~B.}\ \bibnamefont {Agustin}},\
  }\bibfield  {title} {\bibinfo {title} {Evaluation of a journal club designed
  to enhance the professional development of radiation therapists},\
  }\href@noop {} {\bibfield  {journal} {\bibinfo  {journal} {Radiography}\
  }\textbf {\bibinfo {volume} {14}},\ \bibinfo {pages} {120} (\bibinfo {year}
  {2008})}\BibitemShut {NoStop}%
\bibitem [{\citenamefont {Alguirre}(1998)}]{Alguirre1998}%
  \BibitemOpen
  \bibfield  {author} {\bibinfo {author} {\bibfnamefont {P.~C.}\ \bibnamefont
  {Alguirre}},\ }\bibfield  {title} {\bibinfo {title} {A review of journal
  clubs in postgraduate medical education},\ }\href@noop {} {\bibfield
  {journal} {\bibinfo  {journal} {J. Gen. Intern. Med.}\ }\textbf {\bibinfo
  {volume} {13}},\ \bibinfo {pages} {347} (\bibinfo {year} {1998})}\BibitemShut
  {NoStop}%
\bibitem [{\citenamefont {McGlacken-Byrne}\ \emph {et~al.}(2020)\citenamefont
  {McGlacken-Byrne}, \citenamefont {O'Rahelly}, \citenamefont {Cantillon},\
  and\ \citenamefont {Allen}}]{McGlackenByrne2020}%
  \BibitemOpen
  \bibfield  {author} {\bibinfo {author} {\bibfnamefont {S.~M.}\ \bibnamefont
  {McGlacken-Byrne}}, \bibinfo {author} {\bibfnamefont {M.}~\bibnamefont
  {O'Rahelly}}, \bibinfo {author} {\bibfnamefont {P.}~\bibnamefont
  {Cantillon}},\ and\ \bibinfo {author} {\bibfnamefont {N.~M.}\ \bibnamefont
  {Allen}},\ }\bibfield  {title} {\bibinfo {title} {Journal club: {O}ld tricks
  and fresh approaches},\ }\href@noop {} {\bibfield  {journal} {\bibinfo
  {journal} {Arch. Dis. Child. Educ. Pract. Ed.}\ }\textbf {\bibinfo {volume}
  {105}},\ \bibinfo {pages} {236} (\bibinfo {year} {2020})}\BibitemShut
  {NoStop}%
\bibitem [{\citenamefont {Gottlieb}\ \emph {et~al.}(2018)\citenamefont
  {Gottlieb}, \citenamefont {King}, \citenamefont {Byyny}, \citenamefont
  {Parsons},\ and\ \citenamefont {Bailitz}}]{Gottlieb2018}%
  \BibitemOpen
  \bibfield  {author} {\bibinfo {author} {\bibfnamefont {M.}~\bibnamefont
  {Gottlieb}}, \bibinfo {author} {\bibfnamefont {A.}~\bibnamefont {King}},
  \bibinfo {author} {\bibfnamefont {R.}~\bibnamefont {Byyny}}, \bibinfo
  {author} {\bibfnamefont {M.}~\bibnamefont {Parsons}},\ and\ \bibinfo {author}
  {\bibfnamefont {J.}~\bibnamefont {Bailitz}},\ }\bibfield  {title} {\bibinfo
  {title} {Journal club in residency education: {A}n evidence-based guide to
  best practices from the council of emergency medicine residency directors},\
  }\href@noop {} {\bibfield  {journal} {\bibinfo  {journal} {West. J. Emerg.
  Med.}\ }\textbf {\bibinfo {volume} {194}},\ \bibinfo {pages} {746} (\bibinfo
  {year} {2018})}\BibitemShut {NoStop}%
\bibitem [{\citenamefont {Deenadayalan}\ \emph {et~al.}(2008)\citenamefont
  {Deenadayalan}, \citenamefont {Grimmer-Somers}, \citenamefont {Prior},\ and\
  \citenamefont {Kumar}}]{Deenadayalan2008}%
  \BibitemOpen
  \bibfield  {author} {\bibinfo {author} {\bibfnamefont {L.~S.}\ \bibnamefont
  {Deenadayalan}}, \bibinfo {author} {\bibfnamefont {K.}~\bibnamefont
  {Grimmer-Somers}}, \bibinfo {author} {\bibfnamefont {M.}~\bibnamefont
  {Prior}},\ and\ \bibinfo {author} {\bibfnamefont {S.}~\bibnamefont {Kumar}},\
  }\bibfield  {title} {\bibinfo {title} {How to run an effective journal club:
  {A} systematic review},\ }\href@noop {} {\bibfield  {journal} {\bibinfo
  {journal} {J. Eval. Clin. Pract.}\ }\textbf {\bibinfo {volume} {14}},\
  \bibinfo {pages} {898} (\bibinfo {year} {2008})}\BibitemShut {NoStop}%
\bibitem [{\citenamefont {Sanders}\ \emph {et~al.}(2017)\citenamefont
  {Sanders}, \citenamefont {Kohler}, \citenamefont {Faesi}, \citenamefont
  {Villar},\ and\ \citenamefont {Zevin}}]{Sanders2017}%
  \BibitemOpen
  \bibfield  {author} {\bibinfo {author} {\bibfnamefont {N.~H.}\ \bibnamefont
  {Sanders}}, \bibinfo {author} {\bibfnamefont {S.}~\bibnamefont {Kohler}},
  \bibinfo {author} {\bibfnamefont {C.}~\bibnamefont {Faesi}}, \bibinfo
  {author} {\bibfnamefont {A.}~\bibnamefont {Villar}},\ and\ \bibinfo {author}
  {\bibfnamefont {M.}~\bibnamefont {Zevin}},\ }\bibfield  {title} {\bibinfo
  {title} {Incorporating current research into formal higher education settings
  using astrobites},\ }\href@noop {} {\bibfield  {journal} {\bibinfo  {journal}
  {Am. J. Phys.}\ }\textbf {\bibinfo {volume} {85}},\ \bibinfo {pages} {741}
  (\bibinfo {year} {2017})}\BibitemShut {NoStop}%
\bibitem [{\citenamefont {Creswell}\ \emph {et~al.}(2003)\citenamefont
  {Creswell}, \citenamefont {Clark}, \citenamefont {Gutmann},\ and\
  \citenamefont {Hanson}}]{Creswell2003}%
  \BibitemOpen
  \bibfield  {author} {\bibinfo {author} {\bibfnamefont {J.~W.}\ \bibnamefont
  {Creswell}}, \bibinfo {author} {\bibfnamefont {V.~L.~P.}\ \bibnamefont
  {Clark}}, \bibinfo {author} {\bibfnamefont {M.~L.}\ \bibnamefont {Gutmann}},\
  and\ \bibinfo {author} {\bibfnamefont {W.~.}\ \bibnamefont {Hanson}},\
  }\href@noop {} {\emph {\bibinfo {title} {An expanded typology for classifying
  mixed methods research into designs}}}\ (\bibinfo  {publisher} {Thousand
  Oaks, CA: Sage},\ \bibinfo {year} {2003})\BibitemShut {NoStop}%
\bibitem [{\citenamefont {Clark}\ and\ \citenamefont
  {Creswell}(2008)}]{PlanoClark2008}%
  \BibitemOpen
  \bibfield  {author} {\bibinfo {author} {\bibfnamefont {V.~L.~P.}\
  \bibnamefont {Clark}}\ and\ \bibinfo {author} {\bibfnamefont {J.~W.}\
  \bibnamefont {Creswell}},\ }\href@noop {} {\emph {\bibinfo {title} {The mixed
  methods reader}}}\ (\bibinfo  {publisher} {Sage},\ \bibinfo {year}
  {2008})\BibitemShut {NoStop}%
\bibitem [{\citenamefont {Salda\={n}a}(2013)}]{Saldana2013}%
  \BibitemOpen
  \bibfield  {author} {\bibinfo {author} {\bibfnamefont {J.}~\bibnamefont
  {Salda\={n}a}},\ }\href@noop {} {\emph {\bibinfo {title} {The coding manual
  for qualitative researchers}}}\ (\bibinfo  {publisher} {SAGE Publications
  Limited},\ \bibinfo {year} {2013})\BibitemShut {NoStop}%
\bibitem [{\citenamefont {Charmaz}(2006)}]{Charmaz2006}%
  \BibitemOpen
  \bibfield  {author} {\bibinfo {author} {\bibfnamefont {K.}~\bibnamefont
  {Charmaz}},\ }\href@noop {} {\emph {\bibinfo {title} {Constructing grounded
  theory: A practical guide through qualitative analysis}}}\ (\bibinfo
  {publisher} {Sage},\ \bibinfo {year} {2006})\BibitemShut {NoStop}%
\bibitem [{\citenamefont {Kleinpell}(2002)}]{Kleinpell2002}%
  \BibitemOpen
  \bibfield  {author} {\bibinfo {author} {\bibfnamefont {R.~M.}\ \bibnamefont
  {Kleinpell}},\ }\bibfield  {title} {\bibinfo {title} {Rediscovering the value
  of the journal club},\ }\href@noop {} {\bibfield  {journal} {\bibinfo
  {journal} {Am. J. Crit. Care}\ }\textbf {\bibinfo {volume} {11}},\ \bibinfo
  {pages} {412} (\bibinfo {year} {2002})}\BibitemShut {NoStop}%
\bibitem [{\citenamefont {Freshwater}(2011)}]{Freshwater2011}%
  \BibitemOpen
  \bibfield  {author} {\bibinfo {author} {\bibfnamefont {M.~F.}\ \bibnamefont
  {Freshwater}},\ }\bibfield  {title} {\bibinfo {title} {The four {C}’s for a
  journal club – ingredients for success or failure},\ }\href@noop {}
  {\bibfield  {journal} {\bibinfo  {journal} {J. Plast. Reconstr. Aesthet.
  Surg.}\ }\textbf {\bibinfo {volume} {64}},\ \bibinfo {pages} {839} (\bibinfo
  {year} {2011})}\BibitemShut {NoStop}%
\bibitem [{\citenamefont {Weber}\ \emph {et~al.}(2008)\citenamefont {Weber},
  \citenamefont {Maher}, \citenamefont {Powell},\ and\ \citenamefont
  {Lee}}]{Weber2008}%
  \BibitemOpen
  \bibfield  {author} {\bibinfo {author} {\bibfnamefont {K.}~\bibnamefont
  {Weber}}, \bibinfo {author} {\bibfnamefont {C.}~\bibnamefont {Maher}},
  \bibinfo {author} {\bibfnamefont {A.}~\bibnamefont {Powell}},\ and\ \bibinfo
  {author} {\bibfnamefont {H.}~\bibnamefont {Lee}},\ }\bibfield  {title}
  {\bibinfo {title} {Learning opportunities from group discussions: Warrants
  become the objects of debate},\ }\href@noop {} {\bibfield  {journal}
  {\bibinfo  {journal} {Educ. Stud. Math.}\ }\textbf {\bibinfo {volume} {68}},\
  \bibinfo {pages} {247} (\bibinfo {year} {2008})}\BibitemShut {NoStop}%
\bibitem [{\citenamefont {Spillane}\ and\ \citenamefont
  {Crowe}(1998)}]{Spillane1998}%
  \BibitemOpen
  \bibfield  {author} {\bibinfo {author} {\bibfnamefont {A.~J.}\ \bibnamefont
  {Spillane}}\ and\ \bibinfo {author} {\bibfnamefont {P.~J.}\ \bibnamefont
  {Crowe}},\ }\bibfield  {title} {\bibinfo {title} {The role of the journal
  club in surgical training},\ }\href@noop {} {\bibfield  {journal} {\bibinfo
  {journal} {ANZ J. Surg.}\ }\textbf {\bibinfo {volume} {68}},\ \bibinfo
  {pages} {288} (\bibinfo {year} {1998})}\BibitemShut {NoStop}%
\bibitem [{\citenamefont {Stange}\ \emph {et~al.}(2006)\citenamefont {Stange},
  \citenamefont {Miller}, \citenamefont {McLellan}, \citenamefont {Gotler},
  \citenamefont {Phillips}, \citenamefont {Acheson}, \citenamefont {Crabtree},
  \citenamefont {Zyzanski},\ and\ \citenamefont {Nutting}}]{Stange2006}%
  \BibitemOpen
  \bibfield  {author} {\bibinfo {author} {\bibfnamefont {K.~C.}\ \bibnamefont
  {Stange}}, \bibinfo {author} {\bibfnamefont {W.~L.}\ \bibnamefont {Miller}},
  \bibinfo {author} {\bibfnamefont {L.~A.}\ \bibnamefont {McLellan}}, \bibinfo
  {author} {\bibfnamefont {R.~S.}\ \bibnamefont {Gotler}}, \bibinfo {author}
  {\bibfnamefont {W.~R.}\ \bibnamefont {Phillips}}, \bibinfo {author}
  {\bibfnamefont {L.~S.}\ \bibnamefont {Acheson}}, \bibinfo {author}
  {\bibfnamefont {B.~F.}\ \bibnamefont {Crabtree}}, \bibinfo {author}
  {\bibfnamefont {S.~J.}\ \bibnamefont {Zyzanski}},\ and\ \bibinfo {author}
  {\bibfnamefont {P.~S.}\ \bibnamefont {Nutting}},\ }\bibfield  {title}
  {\bibinfo {title} {Annals journal club: It’s time to get radical},\
  }\href@noop {} {\bibfield  {journal} {\bibinfo  {journal} {Ann. Fam. Med.}\
  }\textbf {\bibinfo {volume} {4}},\ \bibinfo {pages} {196} (\bibinfo {year}
  {2006})}\BibitemShut {NoStop}%
\bibitem [{\citenamefont {Harris}\ \emph {et~al.}(2011)\citenamefont {Harris},
  \citenamefont {Kearley}, \citenamefont {C}, \citenamefont {Meats},
  \citenamefont {Roberts}, \citenamefont {Perera},\ and\ \citenamefont
  {Kearley-Shiers}}]{Harris2011}%
  \BibitemOpen
  \bibfield  {author} {\bibinfo {author} {\bibfnamefont {J.}~\bibnamefont
  {Harris}}, \bibinfo {author} {\bibfnamefont {K.}~\bibnamefont {Kearley}},
  \bibinfo {author} {\bibfnamefont {H.}~\bibnamefont {C}}, \bibinfo {author}
  {\bibfnamefont {E.}~\bibnamefont {Meats}}, \bibinfo {author} {\bibfnamefont
  {N.}~\bibnamefont {Roberts}}, \bibinfo {author} {\bibfnamefont
  {R.}~\bibnamefont {Perera}},\ and\ \bibinfo {author} {\bibfnamefont
  {K.}~\bibnamefont {Kearley-Shiers}},\ }\bibfield  {title} {\bibinfo {title}
  {Are journal clubs effective in supporting evidence-based decision making?
  {A} systematic review},\ }\href@noop {} {\bibfield  {journal} {\bibinfo
  {journal} {Med. Teach.}\ }\textbf {\bibinfo {volume} {33}},\ \bibinfo {pages}
  {9} (\bibinfo {year} {2011})},\ \bibinfo {note}
  {https://doi.org/10.3109/0142159X.2011.530321}\BibitemShut {NoStop}%
\bibitem [{\citenamefont {Virtanen}\ \emph {et~al.}(2020)\citenamefont
  {Virtanen}, \citenamefont {Gommers}, \citenamefont {Oliphant}, \citenamefont
  {Haberland}, \citenamefont {Reddy}, \citenamefont {Cournapeau}, \citenamefont
  {Burovski}, \citenamefont {Peterson}, \citenamefont {Weckesser},
  \citenamefont {Bright},\ and\ \citenamefont {et~al.}}]{Virtanen2020}%
  \BibitemOpen
  \bibfield  {author} {\bibinfo {author} {\bibfnamefont {P.}~\bibnamefont
  {Virtanen}}, \bibinfo {author} {\bibfnamefont {R.}~\bibnamefont {Gommers}},
  \bibinfo {author} {\bibfnamefont {T.~E.}\ \bibnamefont {Oliphant}}, \bibinfo
  {author} {\bibfnamefont {M.}~\bibnamefont {Haberland}}, \bibinfo {author}
  {\bibfnamefont {T.}~\bibnamefont {Reddy}}, \bibinfo {author} {\bibfnamefont
  {D.}~\bibnamefont {Cournapeau}}, \bibinfo {author} {\bibfnamefont
  {E.}~\bibnamefont {Burovski}}, \bibinfo {author} {\bibfnamefont
  {P.}~\bibnamefont {Peterson}}, \bibinfo {author} {\bibfnamefont
  {W.}~\bibnamefont {Weckesser}}, \bibinfo {author} {\bibfnamefont
  {J.}~\bibnamefont {Bright}},\ and\ \bibinfo {author} {\bibnamefont
  {et~al.}},\ }\bibfield  {title} {\bibinfo {title} {{SciPy} 1.0: Fundamental
  algorithms for scientific computing in python},\ }\href@noop {} {\bibfield
  {journal} {\bibinfo  {journal} {Nat. Methods}\ }\textbf {\bibinfo {volume}
  {17}},\ \bibinfo {pages} {261} (\bibinfo {year} {2020})}\BibitemShut
  {NoStop}%
\bibitem [{\citenamefont {Fisher}(1956)}]{Fisher1956}%
  \BibitemOpen
  \bibfield  {author} {\bibinfo {author} {\bibfnamefont {R.~A.}\ \bibnamefont
  {Fisher}},\ }\href@noop {} {\emph {\bibinfo {title} {Statistical methods and
  scientific inference}}}\ (\bibinfo  {publisher} {Oxford University Press},\
  \bibinfo {year} {1956})\BibitemShut {NoStop}%
\bibitem [{\citenamefont {Bimczok}\ and\ \citenamefont
  {Graves}(2020)}]{Bimczok2020}%
  \BibitemOpen
  \bibfield  {author} {\bibinfo {author} {\bibfnamefont {D.}~\bibnamefont
  {Bimczok}}\ and\ \bibinfo {author} {\bibfnamefont {J.}~\bibnamefont
  {Graves}},\ }\bibfield  {title} {\bibinfo {title} {A new twist on the
  graduate student journal club: Using a topic‐centered approach to promote
  student engagement},\ }\href@noop {} {\bibfield  {journal} {\bibinfo
  {journal} {Biochem. Mol. Biol. Educ.}\ }\textbf {\bibinfo {volume} {48}},\
  \bibinfo {pages} {262} (\bibinfo {year} {2020})}\BibitemShut {NoStop}%
\bibitem [{\citenamefont {Burris}\ \emph {et~al.}(2019)\citenamefont {Burris},
  \citenamefont {Frederick}, \citenamefont {Malcom}, \citenamefont {Raake},
  \citenamefont {Shin},\ and\ \citenamefont {Daugherty}}]{Burris2019}%
  \BibitemOpen
  \bibfield  {author} {\bibinfo {author} {\bibfnamefont {J.~N.}\ \bibnamefont
  {Burris}}, \bibinfo {author} {\bibfnamefont {E.~K.}\ \bibnamefont
  {Frederick}}, \bibinfo {author} {\bibfnamefont {D.~R.}\ \bibnamefont
  {Malcom}}, \bibinfo {author} {\bibfnamefont {S.}~\bibnamefont {Raake}},
  \bibinfo {author} {\bibfnamefont {M.}~\bibnamefont {Shin}},\ and\ \bibinfo
  {author} {\bibfnamefont {K.~K.}\ \bibnamefont {Daugherty}},\ }\bibfield
  {title} {\bibinfo {title} {Impact of a journal club elective course on
  student learning measures},\ }\href@noop {} {\bibfield  {journal} {\bibinfo
  {journal} {Am. \ J. \ Pharm.\ Educ.}\ }\textbf {\bibinfo {volume} {83}}
  (\bibinfo {year} {2019})}\BibitemShut {NoStop}%
\bibitem [{\citenamefont {G.Lee}\ \emph {et~al.}(2005)\citenamefont {G.Lee},
  \citenamefont {Boldt}, \citenamefont {Golnik}, \citenamefont {C.Arnold},
  \citenamefont {Oetting}, \citenamefont {Beaver}, \citenamefont {J.Olson},\
  and\ \citenamefont {Carter}}]{Lee2005}%
  \BibitemOpen
  \bibfield  {author} {\bibinfo {author} {\bibfnamefont {A.}~\bibnamefont
  {G.Lee}}, \bibinfo {author} {\bibfnamefont {H.~C.}\ \bibnamefont {Boldt}},
  \bibinfo {author} {\bibfnamefont {K.~C.}\ \bibnamefont {Golnik}}, \bibinfo
  {author} {\bibfnamefont {A.}~\bibnamefont {C.Arnold}}, \bibinfo {author}
  {\bibfnamefont {T.~A.}\ \bibnamefont {Oetting}}, \bibinfo {author}
  {\bibfnamefont {H.~A.}\ \bibnamefont {Beaver}}, \bibinfo {author}
  {\bibfnamefont {R.}~\bibnamefont {J.Olson}},\ and\ \bibinfo {author}
  {\bibfnamefont {K.}~\bibnamefont {Carter}},\ }\bibfield  {title} {\bibinfo
  {title} {Using the journal club to teach and assess competence in
  practice-based learning and improvement: a literature review and
  recommendation for implementation},\ }\href@noop {} {\bibfield  {journal}
  {\bibinfo  {journal} {Surv. Ophthalmol.}\ }\textbf {\bibinfo {volume} {50}},\
  \bibinfo {pages} {542} (\bibinfo {year} {2005})}\BibitemShut {NoStop}%
\bibitem [{\citenamefont {Brown}(999b)}]{Brown1999b}%
  \BibitemOpen
  \bibfield  {author} {\bibinfo {author} {\bibfnamefont {C.~M.}\ \bibnamefont
  {Brown}},\ }\bibfield  {title} {\bibinfo {title} {Information‐seeking
  behavior of scientists in the electronic information age: astronomers,
  chemists, mathematicians, and physicists},\ }\href@noop {} {\bibfield
  {journal} {\bibinfo  {journal} {J. Am. Soc. Inform. Sci.}\ }\textbf {\bibinfo
  {volume} {50}},\ \bibinfo {pages} {929} (\bibinfo {year}
  {1999b})}\BibitemShut {NoStop}%
\end{thebibliography}
